\documentclass{article}
\usepackage[a4paper,margin=2.5cm]{geometry}
\usepackage[utf8]{inputenc}

\usepackage[style=apa,backend=biber]{biblatex}
\usepackage{amssymb, amsmath, amsthm, mathtools}
\usepackage{times}
\usepackage{bm, graphicx, xcolor, subcaption, booktabs}
\usepackage{multirow}
\usepackage[colorlinks=true,
            citecolor = blue,
            linkcolor = blue,
            urlcolor=blue]{hyperref} 
\usepackage{enumitem} 
\usepackage{threeparttable}
\usepackage{orcidlink}

\usepackage{minted}

\newcommand{\ind}{\mathbb{I}}

\newtheorem{proposition}{Proposition}[section]

\newtheorem{remark}{Remark}[section]

\title{Bayesian structured additive quantile regression for inflated bounded data}

\author{Francisco F. Queiroz\,\orcidlink{0000-0001-8368-0707}$^1$\footnote{Corresponding author: Francisco F. Queiroz, email felipeq@ime.usp.br.}, ~Johannes Brachem\,\orcidlink{0000-0001-7884-4631}$^2$,\\ Paul F.V.\ Wiemann\,\orcidlink{0000-0003-1901-0295}$^3$, ~ Thomas Kneib\,\orcidlink{0000-0003-3390-0972}$^2$,\\ \\
{\small {\em $^1$Department of Statistics, University of S\~ao Paulo, Brazil}}\\
{\small {\em $^2$Chair of Statistics, University of Göttingen, Germany}}\\
{\small {\em $^2$Department of Statistics, The Ohio State University, United States}}}
\date{}

\begin{document}
\maketitle
\begin{abstract}
\noindent Bounded continuous data on the unit interval frequently arise in applied fields and often exhibit a non-negligible proportion of observations at the boundaries. Inflated regression models address this feature by combining a continuous distribution on the unit interval with a discrete component to account for zero- and/or one-inflation. In this paper, we propose a class of Bayesian structured additive quantile regression models for inflated bounded continuous data that accommodates zero- and/or one-inflation. The proposed approach enables direct modeling of both the conditional quantiles of the continuous component and the probabilities of observing zeros and/or ones, with structured additive predictors incorporated in both parts, including nonlinear effects, spatial effects, random effects, and varying-coefficient terms. Posterior inference is carried out using Markov chain Monte Carlo algorithms implemented through the software Liesel, a probabilistic programming framework for semiparametric regression. The practical performance of the proposed models is illustrated through simulation studies and two real-data applications: one analyzing the proportion of traffic-related fatalities across Brazilian municipal districts, and another evaluating speech intelligibility in cochlear implant recipients under different experimental conditions.

\end{abstract}

\section{Introduction}\label{sec:intro}

Bounded continuous data on the unit interval are common in fields such as medicine, biology, sociology, and ecology, with examples including mortality rates, body fat percentages, and the proportion of land covered by tree canopies. Oftentimes, these data contain a non-negligible amount of observations at the boundaries, i.e., significant numbers of zeros and/or ones. As examples, consider the mortality or infection rate for a certain disease, the proportion of family income spent on children's education, and the proportion of people who have a certain characteristic. A natural approach to modeling this type of data is to use an inflated regression model, where the response variable $y$ follows a mixture distribution composed of a continuous random variable on $(0,1)$  and a discrete random variable which accounts for boundary values. The probability density function (PDF) of a zero- and one-inflated random variable can generally be expressed as
\begin{align}\label{eq:pdfbeta}
    g(y; p_0, p_1, \bm{\vartheta}) = \left\{ p_0^{\ind_{\{0\}}(y)} p_1^{\ind_{\{1\}}(y)} (1-p_0-p_1)^{\ind_{(0,1)}(y)}  \right\} \left\{ h(y; \bm{\vartheta})^{\ind_{(0,1)}(y)}  \right\},\quad y \in [0,1],
\end{align}
where $\ind_{A}(y)$ is the indicator function that is equal to $1$ if $y \in A$ and $0$ otherwise, $h(y; \bm{\vartheta})$ represents the PDF of the continuous random variable that may depend on a parameter vector $\bm{\vartheta}$, $0<p_0, p_1<1$ and $p_0+p_1<1$. The parameters $p_0$ and $p_1$ represent the probabilities of zero and one, respectively. The PDF of a zero- or one-inflated random variable can be derived from Equation~\eqref{eq:pdfbeta} by setting $p_1=0$ for zero-inflation or $p_0=0$ for one-inflation. Note that the PDF $g$ factorizes into two terms: the first one depends only on $p_0$ and $p_1$ (the parameters of the discrete part of the mixture) and the second one only on $\bm{\vartheta}$ (the parameters of the continuous part of the mixture). Thus, we say that the parameters $(p_0, p_1)$ and $\bm{\vartheta}$ are separable and inference based on the likelihood can be performed independently for the discrete and the continuous parts. Additionally, note that the first term in Equation~\eqref{eq:pdfbeta} corresponds to the probability function of a multinomial random variable with a single trial and a vector of probabilities $(p_0, p_1, 1-p_0-p_1)$ when we consider the variable $(\ind_{\{0\}}(y), \ind_{\{1\}}(y), \ind_{(0,1)}(y))$. This term reduces to the probability function of a Bernoulli distribution if either $p_0$ or $p_1$ is equal to zero.

Several models have been proposed in the literature for modeling inflated bounded continuous data, where the response variable takes the form given in Equation~\eqref{eq:pdfbeta}. Specifically, if $h(y; \bm{\vartheta})$ is the PDF of a given continuous distribution defined on $(0,1)$, a corresponding zero- and/or one-inflated regression model for bounded data is formulated. Examples include models based on the beta distribution \parencite{ospinaFerrari, ospinaFerrari2010, stasinopoulos2024gamlss}, the unit-Weibull distribution \parencite{Menezes2021}, and the power logit distribution \parencite{Queiroz2024}, among others. In all these models, both likelihood-based and Bayesian inference are typically carried out separately for the parameters of the discrete and continuous components, due to the separability of the parameter vectors and, consequently, of the likelihood function. We will refer to this modeling strategy, which treats the two components independently, as the two-part modeling framework. Another approach to deal with boundary observations was recently proposed by \textcite{Kosmidis2025}. The authors introduced the extended-support beta (XBX) regression model, which extends the standard beta regression to allow positive probability at the boundaries. This approach is particularly appealing in situations where it is reasonable to assume that both boundary and continuous observations arise from a single underlying data-generating process, rather than from distinct discrete and continuous mechanisms.

Many regression models characterize the conditional distribution of the response through a limited set of parameters, often targeting one or two specific aspects such as the mean and precision. While such parameter-focused models can be very successful, they allow covariate effects only on limited aspects of the conditional distribution and introduce the assumption of a parametric distribution. A more flexible perspective is offered by quantile regression, which directly targets conditional quantiles and thus enables a more comprehensive analysis that captures heterogeneity, asymmetry, and covariate effects beyond average trends without a distributional assumption. Quantile regression was proposed by \textcite{Koenker1978} and has been applied in various contexts; see, for instance, \textcite{Koenker2004}, \textcite{Kneib2013}, \textcite{Razen2023} and references therein. Some studies in the literature have explored quantile regression methods for modeling bounded data in the interval $(0,1)$. For example, \textcite{Bottai2010} introduced quantile regression for bounded continuous data based on the logistic transformation, while \textcite{bayes2017} proposed a parametric quantile mixed regression approach using the Kumaraswamy distribution.  More recently, \textcite{Bourguignon2024} proposed a parametric quantile regression model based on a generalized three-parameter beta distribution. 

There are few studies that deal with quantile regression in the context of inflated regression models for bounded data. Some examples are presented in \textcite{Santos2015} and in \textcite{Menezes2021}, where quantile regression models are proposed for handling zero- or one-inflated bounded continuous data. The approach employed by \textcite{Menezes2021} parallels that of \textcite{Bourguignon2024}, utilizing a quantile-based parameterization within the unit-Weibull distribution to develop a quantile regression for the continuous part of the model. The approach presented by \textcite{Santos2015} closely resembles the concept of quantile regression initially introduced by \textcite{Koenker1978}. The authors propose a Bayesian quantile regression model to explain the conditional distribution of the continuous part, while the mixture probability is modeled as a function of the covariates. They utilize the asymmetric Laplace distribution as an auxiliary error distribution. 

The models proposed by \textcite{Santos2015} and \textcite{Menezes2021} employ simple linear regression specifications for the predictors and are not applicable when the data contain observations at both boundaries, that is, in the presence of zero- and one-inflation. In this paper, we introduce more flexible types of quantile regression models for inflated bounded continuous data, with their parameters estimated under a Bayesian framework. These models provide a flexible approach for capturing the distributional features of inflated bounded data.

The main contributions of this work are twofold. First, the proposed models incorporate structured additive regression, allowing for the inclusion of various structured effects, such as smooth functions of covariates, spatial effects, random effects, and varying-coefficient terms. Second, the models are designed to handle both zero-or-one and zero- and one-inflated bounded data. To the best of our knowledge, no existing quantile regression models currently address zero- and one-inflated data. Furthermore, the methodology developed in this work provides a flexible foundation that can be extended to other classes of inflated models.

The remainder of this paper is organized as follows. Section \ref{sec:boundedquantreg} introduces the Bayesian structured additive quantile regression models for zero- and/or one-inflated bounded continuous data. It also discusses the two-part modeling framework, the structured additive predictors, and Bayesian inference via the asymmetric Laplace distribution. Section \ref{sec:computationalimplementation} describes the posterior sampling scheme. Simulation studies are presented in Section \ref{sec:numerical-results}. Finally, Section \ref{sec:applications} illustrates the practical use of the proposed models in two real-data applications involving traffic-related fatalities and speech intelligibility experiments. The article closes with a discussion of the main findings and directions for future research.

\section{Bayesian structured additive quantile regression models for inflated bounded data}\label{sec:boundedquantreg}

We propose a Bayesian semiparametric structured additive quantile regression model for zero- and/or one-inflated data, formulated within the two-part modeling framework. The first part addresses the discrete component by modeling the probability of inflation at a boundary point (either zero, one, or both), while the second part focuses on the continuous component, capturing the conditional quantiles of the bounded response variable. Without loss of generality, we assume that the response variable is defined on the unit interval, as the proposed approach can be readily extended to any bounded interval with known boundaries.

In order to formalize the zero- and one-inflated structured additive quantile regression models, let $(y_i, \bm{x}_i)$, $i=1, \ldots, n$, denote the observations of the response variable $y_i$ and the corresponding covariate vector $\bm{x}_i=(x_{1i}, ..., x_{qi})^\top$ for the $i$th individual. We assume that the responses are conditionally independent given the covariates. The response variable $y_i$ takes values in the set $[0,1]$, allowing for inflation at both boundaries, zero and one. Let $p_{0i} = \mathbb{P}(y_i = 0)$ and $p_{1i} = \mathbb{P}(y_i = 1)$ denote the probabilities that the response takes the values zero and one, respectively, with $p_{0i} \in (0,1)$, $p_{1i} \in (0,1)$, and $p_{0i} + p_{1i} <1$, for all $i=1, \ldots, n$. Let $Q_{y_i|y_i \in (0,1)} (\tau | \bm{x}_i)$, for $i=1, \ldots, n$, denote the $\tau$th conditional quantile of $y_i$ given $\bm{x}_i$, restricted to observations in the interval $(0,1)$. Since our response variable is restricted to a bounded interval, instead of modeling directly $Q_{y_i|y_i \in (0,1)} (\tau | \bm{x}_i)$, we model $Q_{h(y_i)|y_i \in (0,1)} (\tau | \bm{x}_i)$, where $h: (0,1) \rightarrow \mathbb{R}$ is a strictly increasing link function. We rely on the fact that
\[
    Q_{y_i|y_i \in (0,1)} (\tau | \bm{x}_i) = h^{-1}\left[Q_{h(y_i)|y_i \in (0,1)} (\tau | \bm{x}_i) \right]
\]
to obtain the conditional quantiles of the response variable on the original scale. This approach allows us to model the conditional quantile on the real line using standard quantile regression techniques. Rather than assuming simple linear predictors for the inflation probability $p_i$ and the conditional quantile $Q_{y_i|y_i \in (0,1)} (\tau | \bm{x}_i)$, we adopt a flexible semiparametric regression framework with structured additive predictors \parencite{Waldmann2013, Kneib2023, fahrmeir2021regression} of the generic form
\begin{align}
    d_0\left(\dfrac{p_{0i}}{p_{2i}}\right) = \eta_{0i} & =  \beta_{00} +  \sum_{k=1}^{m_0} f_{0k}(\bm{x}_i), \label{eq:disc1} \\
    d_1\left(\dfrac{p_{1i}}{p_{2i}}\right)= \eta_{1i} & =  \beta_{10} +  \sum_{k=1}^{m_1} f_{1k}(\bm{x}_i),  \label{eq:disc2} \\
    h\left[Q_{y_i|y_i \in (0,1)} (\tau | \bm{x}_i) \right] = \eta_{2i, \tau} & =  \beta_{20, \tau} +  \sum_{k=1}^{m_2} f_{2k, \tau}(\bm{x}_i),\label{eq:continuous}
\end{align}
where $d_0, d_1: (0,1) \rightarrow \mathbb{R}$ are strictly monotonic and
twice differentiable link functions, $\beta_{00}$, $\beta_{10}$ and $\beta_{20, \tau}$ are intercepts, and $f_{0k}(\bm{x}_i)$, $f_{1k}(\bm{x}_i)$ and $f_{2k, \tau}(\bm{x}_i)$ denote generic functions capturing different types of regression effects, each depending on (subsets of) the full covariate vector $\bm{x}_i$.
A natural choice for both $d_0$ and $d_1$ is the logarithm link function corresponding to a multinomial logit parameterization and for $h$ is the logit link, defined as $\text{logit}(x) = \log[x/(1-x)]$, $x \in (0,1)$.
For notational simplicity, the dependence on $\tau$ in $\beta_{20, \tau}$, $\eta_{2i, \tau}$, and in the functions $f_{2k, \tau}(\bm{x}_i)$ is omitted in the following descriptions, and we simply write $\beta_{20}$, $\eta_{2i}$, and $f_{2k}(\bm{x}_i)$, respectively. 

Some examples of common effects are
\begin{itemize}
    \item[i.] linear: $f(\bm{x}_i) = \bm{x}_i^\top \bm{\beta}$, where $\bm{\beta}$ is the parameter vector, which depends on the $\tau$th quantile;
    \item[ii.] smooth  semi-parametric component: $f(\bm{x}_i) = f_{\text{smooth}}(x_{ri})$, where $x_{ri}$ is a continuous covariate of $\bm{x}_i$ and $f_{\text{smooth}}$ is a smooth function that satisfies certain conditions;
    \item[iii.] spatial component: $f(\bm{x}_i) = f_{\text{spatial}}(\bm{s}_i)$, where $\bm{s}_i$ is a vector of spatial location variable;
    \item[iv.] varying coefficients term: $f(\bm{x}_i) = x_{ri}f_{\text{smooth}}(x_{si})$, where $x_{ri}$ is a continuous or binary covariate and its effect is modified by $x_{si}$ (continuous covariate). 
\end{itemize}
Each of the nonlinear effects in the structured additive predictors from Equations~\eqref{eq:disc1}-\eqref{eq:continuous} is represented via basis function expansions as
\begin{align*}
    f_{jk}(\bm{x}_i) = \sum_{l = 1}^{L_{jk}} \beta_{jkl} B_{jkl}(\bm{x}_i),
\end{align*}
for $j=0,1,2$, where $B_{jkl}(\bm{x}_i)$, for $l=1, \ldots, L_{jk}$, are the basis functions that depends on the specific effect type and $\beta_{jkl}$ are their corresponding coefficients. In matrix notation, the predictors can be written as
\begin{align}\label{eq:predictors-matrix}
    \bm{\eta}_{j} & =  \beta_{j0}\bm{1}_n + \bm{B}_{j1} \bm{\beta}_{j1} + \cdots  + \bm{B}_{j{m_j}} \bm{\beta}_{j{m_j}},
\end{align}
for $j=0,1,2$, where $\bm{\beta}_{jk} = (\beta_{jk1}, \ldots, \beta_{jkL_{jk}} )^\top$ is the vector of basis coefficients and $\bm{B}_{jk}$ are the design matrices obtained from the basis functions evaluated at the covariates, i.e., $\bm{B}_{jk}[i, l] =  B_{jkl}(\bm{x}_i)$. See \textcite[Chaps.~8--9]{fahrmeir2021regression} for details.

In the two-part modeling framework, the discrete component is specified through a multinomial regression model, where the dependent variable, namely $\left(\mathbb{I}_{\{0\}}(y_i), \mathbb{I}_{\{1\}}(y_i), \mathbb{I}_{(0,1)}(y_i)\right)$, $i = 1, \ldots, n$, is assumed to be conditionally independent given the covariates and follows a Multinomial distribution with a single trial and a vector of probabilities $(p_{0i}, p_{1i}, p_{2i})$. The continuous component models the conditional quantiles of the transformed response variable, $h(y_i)$, for observations in the set $\mathcal{I} = \{ i : y_i \in (0,1) \}$. Without assuming a full distribution for $h(y_i)$, estimation in quantile regression may be performed by minimizing the sum of asymmetrically weighted absolute deviations (AWADs), typically via linear programming techniques; see \textcite{koenker2005quantile} for details. For Bayesian inference, however, the asymmetric Laplace distribution (ALD) is commonly employed as a working model due to its convenient location-scale mixture representation, which substantially simplifies posterior computation \parencite{ kozumi2011gibbs, Waldmann2013, Santos2015}. Therefore, we assume that $y_i^\dagger \coloneqq h(y_i)$ follows the ALD with location parameter $\eta_{2i}$ as specified in Equation~\eqref{eq:continuous}, precision parameter $\delta^2$, skewness parameter $\tau$, and PDF given by
\[
    g_{y_i^\dagger}(y; \eta_{2i}, \delta^2, \tau) = \tau (1-\tau)\delta^2 \exp\{- \delta^2 \rho_\tau (y- \eta_{2i})  \}, \quad y \in \mathbb{R},
\]
where $\rho_\tau(v) = \tau |v|$ if $v \geq 0$ and $\rho_\tau(v) =(1- \tau) |v|$ otherwise. Under this assumption, the $\tau$th quantile of $h(y_i)$ is $\eta_{2i}$. The posterior modes obtained with the ALD coincide with the minimizers of the sum of AWADs; see \textcite{Waldmann2013}. 

We consider the location-scale mixture representation of the ALD proposed by \textcite{yue2011bayesian}. 
\begin{proposition}[Location-scale mixture representation of the ALD]
Let $z \sim \text{N}(0,1)$ and  $w \sim \text{Exp}(\delta^2)$ be two independent random variables following a standard normal and exponential distribution with rate parameter $\delta^2$, respectively. Then, 
\[
    y^\dagger = \eta + \xi w + \sigma z \sqrt{\dfrac{w}{\delta^2}} \sim \text{ALD}(\eta, \delta^2, \tau),
\]
where $\xi = (1-2\tau)/[\tau(1-\tau)]$ and $\sigma^2 = 2/[\tau(1-\tau)]$.
\end{proposition}
\noindent Thus, once $w$ is imputed, the model for Bayesian inference becomes a conditionally Gaussian regression with offsets $\xi w$ and weights $\sigma \sqrt{w/\delta^2}$.

Based on the two-part modeling framework, we have the following specifications for our observed model:
\[
    \left(\mathbb{I}_{\{0\}}(y_i), \mathbb{I}_{\{1\}}(y_i), \mathbb{I}_{(0,1)}(y_i)\right)|\beta_{00}, \bm{\beta}_{01}, \ldots, \bm{\beta}_{1m_0}, \beta_{10}, \bm{\beta}_{11}, \ldots, \bm{\beta}_{1m_1}  \sim \text{Multinomial}(1; p_{0i}, p_{1i}, p_{2i})  ,
\]
for $i = 1, \ldots, n$, and
\[
    h(y_i)|\beta_{20}, \bm{\beta}_{21}, \ldots, \bm{\beta}_{2m_2}, w_i, \delta^2  \sim \text{N}(\eta_{2i} + \xi w_i, w_i\sigma^2/\delta^2  )  , \quad i \in \mathcal{I},
\]
where $w_i|\delta^2  \sim \text{Exp}(\delta^2)$. We assume a gamma distribution, $\text{Gamma}(a_0, b_0)$, with $a_0=b_0 = 0.001$, as the prior for $\delta^2$. In order to impose specific properties on the estimated functions in Equations~\eqref{eq:disc1}-\eqref{eq:continuous}, each coefficient vector $\bm{\beta}_{jk}$ is assigned a potential rank-deficient multivariate normal prior distribution with density:
\begin{align}\label{eq:priorbeta}
    \pi(\bm{\beta}_{jk} | \nu^2_{jk}) \propto \left( \dfrac{1}{\nu^2_{jk}} \right)^{\text{rank}(\textbf{K}_{jk})/2} \exp \left\{ - \dfrac{1}{2\nu_{jk}^2} \bm{\beta}_{jk}^\top \textbf{K}_{jk} \bm{\beta}_{jk} \right\},
\end{align}
where $j=0,1,2$, $k=1, \ldots, m_j$, $\textbf{K}_{jk}$ is a positive-semidefinite and potentially rank-deficient penalty matrix chosen to induce the desired regularization properties in the effects, and $\nu^2_{jk}$ is a prior smoothing variance that determines the influence of the prior on the posterior. An inverse gamma hyperprior is assumed for $\nu^2_{jk}$, that is, $\nu^2_{jk} \sim \mathrm{IG}(a_{jk}, b_{jk})$, with weakly informative choices $a_{jk} = b_{jk} = 0.01$. The prior specification for $\bm{\beta}_{jk}$ can be directly related to the penalized likelihood framework, in which each effect is subject to a quadratic penalty of the form $\text{pen}(\bm{\beta}_{jk}) = \lambda_{jk} \bm{\beta}_{jk}^\top \textbf{K}_{jk} \bm{\beta}_{jk}$. Here, $\lambda_{jk}$ a smoothing parameter that plays a role analogous to that of $\nu^{-2}_{jk}$. Details on how to construct design and penalty matrices for different types of effects are provided in \textcite[Chap.~9]{fahrmeir2021regression}; see also \textcite{Waldmann2013} and \textcite{Klein2015}.

Assuming that all priors are mutually independent, the unnormalized posterior is given by
\begin{align*}
    \pi(\bm{\theta}| \bm{y} ) & \propto \left\{ \prod_{i=1}^n \left[  p_{0i}^{\ind_{\{0\}}(y_i)}  p_{1i}^{\ind_{\{1\}}(y_i)} p_{2i}^{\ind_{(0,1)}(y_i)} \right] \right\} \left[ \prod_{k=1}^{m_0} \pi(\bm{\beta}_{0k} | \nu^2_{0k}) \pi(\nu^2_{0k}) \right] \left[ \prod_{k=1}^{m_1} \pi(\bm{\beta}_{1k} | \nu^2_{1k}) \pi(\nu^2_{1k}) \right] \\
    & \hspace{1.5em} \times  \left[ \prod_{i \in \mathcal{I}} \phi(y_i^\dagger;\eta_{2i} + \xi w_i, w_i\sigma^2/\delta^2 ) \pi(w_i|\delta^2) \right]  \left[ \prod_{k=1}^{m_2} \pi(\bm{\beta}_{2k} | \nu^2_{2k}) \pi(\nu^2_{2k}) \right] \pi(\delta^2),
\end{align*}
where $\bm{\theta}$ collects all model parameters, $\bm{y} = (y_1, \ldots, y_n)^\top$, and $\phi(\cdot; a, b)$ is the PDF of a normal distribution with mean $a$ and variance $b$.

\paragraph{One-sided inflation}
The zero- or one-inflated structured additive quantile regression models are limiting cases of the zero- and one-inflated structured additive quantile regression models, obtained when $p_{0i} \to 0$ (yielding one-inflation) or $p_{1i} \to 0$ (yielding zero-inflation). In this setting, the response variable $y_i$ takes value in the set $\{c\} \cup (0,1)$, where $c=0$ or $c=1$ depending on the case, and the discrete component is specified through a binary regression model, where the dependent variable, namely $\mathbb{I}_{\{c\}}(y_i)$, $i = 1, \ldots, n$, is assumed to be conditionally independent given the covariates and follows a Bernoulli distribution with success probability $p_{i} = \mathbb{P}(y_i=c)$. For zero-inflation, we set $p_{i} = p_{0i}$ and the model predictors are given by Equations~\eqref{eq:disc1}-\eqref{eq:continuous}, with $p_{2i}=1-p_{0i}$. Analogously, for one-inflation, we set $p_{i} = p_{1i}$ and the model predictors are given by Equations~\eqref{eq:disc2}-\eqref{eq:continuous}, with $p_{2i}=1-p_{1i}$. Thus, when the logarithmic link function is considered in the discrete part, the resulting discrete model corresponds to a structured additive logistic regression model, with predictor given by $\text{logit}(p_{i})$. The remaining model specifications, including prior distributions and inference procedures, follow analogously from the general zero- and one-inflated formulation. For the sake of completeness, the unnormalized posterior of $\bm{\theta}$ for the zero- or one-inflated structured additive quantile regression models is given by
\begin{align*}
    \pi(\bm{\theta}| \bm{y} ) & \propto \left\{ \prod_{i=1}^n \left[  p_i^{\ind_{\{c\}}(y_i)}  (1-p_i)^{1 - \ind_{\{c\}}(y_i)} \right] \right\} \left[ \prod_{k=1}^{m_c} \pi(\bm{\beta}_{ck} | \nu^2_{ck}) \pi(\nu^2_{ck}) \right] \\
    & \hspace{1.5em} \times  \left[ \prod_{i \in \mathcal{I}} \phi(y_i^\dagger;\eta_{2i} + \xi w_i, w_i\sigma^2/\delta^2 ) \pi(w_i|\delta^2) \right]  \left[ \prod_{k=1}^{m_2} \pi(\bm{\beta}_{2k} | \nu^2_{2k}) \pi(\nu^2_{2k}) \right] \pi(\delta^2),
\end{align*}
where $\bm{\theta}$ collects all model parameters.

\begin{remark}
    If all functions in the predictors are linear, we obtain the zero- or one-inflated quantile regression models proposed by \textcite{Santos2015}.
\end{remark}

We conclude this section with two remarks regarding the proposed modeling framework. First, although the joint posterior distributions are specified for all model parameters, the parameters associated with the discrete component do not depend on the quantile level $\tau$ and can therefore be estimated once and used across all quantile levels. In contrast, the parameters of the continuous component are quantile-specific and must be estimated separately for each $\tau$. Second, since the continuous component is defined only for observations in $(0,1)$, its estimated effects should be interpreted as describing the conditional quantiles of the response given $y_i \in (0,1)$.

\section{Posterior sampling}\label{sec:computationalimplementation}

We rely on Markov chain Monte Carlo (MCMC) methods to draw samples from the posterior distributions of the parameters of the zero- and one-inflated structured additive quantile regression models. Samples are drawn blockwise by iteratively sampling from the full conditional posterior distributions.
For the regression coefficients, we use Metropolis-Hastings steps with tuned iteratively weighted least squares proposals similar to \textcite{Klein2015}; see also \textcite{Gamerman1997}. The proposal distribution is a multivariate normal, 
\begin{align}
 \boldsymbol{\beta}_{jk}^* \sim N(\mathbf{b}_{jk}, \epsilon_{jk}^2 \mathbf{F}(\boldsymbol{\beta}_{jk})^{-1}), \quad \text{with} \quad
 \mathbf{b}_{jk} = \boldsymbol{\beta}_{jk} + [\epsilon_{jk}^2/2]\mathbf{F}(\boldsymbol{\beta}_{jk})\mathbf{s}(\boldsymbol{\beta}_{jk}),
\end{align}
where $\mathbf{F}(\boldsymbol{\beta}_{jk})$ and $\mathbf{s}(\boldsymbol{\beta}_{jk})$ denote the negative Hessian and the gradient of the log full conditional posterior evaluated at $\boldsymbol{\beta}_{jk}$, respectively, and $\epsilon_{jk}$ is a step size that is tuned to reach an acceptance probability of $0.8$ using the dual averaging algorithm  \parencite{Hoffman2014-NoUTurnSamplerAdaptively} in a warmup phase. The negative Hessian and gradient are evaluated using automatic differentiation via JAX \parencite{deepmind2020jax}.

The smoothing variances $\nu_{jk}^2$ are sampled using Gibbs updates, drawing from the inverse Gamma full conditional distributions 

$$
\nu_{jk}^2 \mid \cdot \sim \mathrm{IG}\bigl(a_{jk} + 0.5 \mathrm{rank}(\mathbf{K}_{jk}),\ b_{jk} + 0.5 \boldsymbol{\beta}_{jk}^\top\mathbf{K}_{jk}\boldsymbol{\beta}_{jk}\bigr).
$$
To obtain samples for $w_i$, we use the location-scale mixture representation of the ALD, which implies that the $w_i$ are independent and identically distributed given $\delta^2$, with  $w_i|\delta^2 \sim \text{Exp}(\delta^2)$, for $i \in \mathcal{I}$. Consequently, the full conditional distributions for $w_i^{-1}$ are inverse Gaussian, that is, 
\[
    w_i^{-1}| \cdot \sim \text{InvGauss} \left( \sqrt{\dfrac{\xi^2 + 2\sigma^2}{(y_i^\dagger - \eta_{2i})^2}} , \dfrac{\delta^2(\xi^2 + 2\sigma^2)}{\sigma^2} \right), \quad i \in \mathcal{I}.
\]
See \textcite{Waldmann2013} for further details. We draw Gibbs updates for $w_i^{-1}$ from this full conditional. The remaining model parameters, $w_1, \dots, w_n$, and $\delta$, are updated on log-level using general No-U-Turn Samplers \parencite[NUTS,][]{Hoffman2014-NoUTurnSamplerAdaptively}.

The models are implemented in Liesel, a probabilistic programming framework designed for semi-parametric regression \parencite{Riebl2023}. The basis and penalty matrices for covariate effects are obtained internally from the \texttt{mgcv} package in R \parencite{Woodmgcv, Woodbook}. Further details regarding the implementation of the proposed models are presented in Section 1 of the Supplementary Material. The complete code illustrating the implementation of the proposed models can be found at the GitHub repository \href{https://github.com/ffqueiroz/inflated_quantreg}{inflated\_quantreg}.

\section{Numerical results}\label{sec:numerical-results}

We conduct a simulation study to evaluate the proposed implementation of the structured additive quantile regression for inflated bounded data, as well as the impact of likelihood misspecification when using the ALD to model the continuous component. Here, we present the results considering zero- and one-inflation, although similar findings are observed for zero- or one-inflation.

We generate the data from a zero- and one-inflated regression model, with the following specifications for the discrete component:
\begin{align*}\label{linearpred}
    \begin{split}
        \log\left( \dfrac{p_{0i}}{p_{2i}} \right) &= 0.3 \times [m_1(x_{1i}) + m_2(x_{2i})] - k,\\
        \log\left( \dfrac{p_{1i}}{p_{2i}} \right) &= 0.3 \times [m_3(x_{1i}) + m_4(x_{2i})] - k,
    \end{split} 
\end{align*}
 for $i=1, \ldots, n$, where $m_1(x) = \sin(3 \pi x)e^{-x}$, $m_2(x) = x^3$, $m_3(x) = 0.5\exp(-x^2)-0.2$, and $m_4(x) = 1$, $p_{2i} = 1-p_{0i}-p_{2i}$, and the constant $k$ is selected to generate two ranges for the probability of observing $\{0,1\}$: small $(p_{0i}+p_{1i} \in (0.15,\,0.21),\, k=2.5)$ and moderate $(p_{0i}+p_{1i} \in (0.41,\,0.58),\, k=1)$. We consider two scenarios to generate the data for the continuous part:
 \begin{enumerate}
     \item[S1.]  we obtain the continuous data by applying the logit transformation to data drawn from the ALD with location parameter $\eta_{2i}$, precision parameter $\delta^2 = 9$, and skewness parameter $\tau \in \{0.1, 0.5, 0.9\}$, where 
    $$\eta_{2i}  = m_1(x_{1i}) + m_2(x_{2i}).$$
    Note that $\eta_{2i}$ corresponds to the $\tau$th quantile of the logit-transformed continuous response variable. Thus, since we are using the ALD to estimate the quantiles of the continuous part, this is a non-mispecified scenario. 
    \item[S2.] we generate the continuous data from the GJS-t$_{(\varrho)}$ distribution \parencite{LemonteBazan2016} with median $\mu_i$, precision $\sigma_i = x_{2i}$, and degrees-of-freedom $\varrho = 4$, where
        $$\text{logit}(\mu_i)  = m_1(x_{1i}) + m_2(x_{2i}).$$
    The $\tau$th quantile of the GJS-t$_{(\varrho)}$ distribution is given by
    \begin{equation}\label{eq:quantileGJSt}
        y^\tau_i =  \dfrac{\mu_i e^{\sigma_i t^\tau}}{1 -  \mu_i (1 -e^{\sigma_i t^\tau})},
    \end{equation} 
    where $t^\tau$ is the $\tau$th quantile of a Student-t distribution with four degrees of freedom.
 \end{enumerate}
The covariates $x_{1i}$ and $x_{2i}$ are independently drawn from a uniform distribution on the unit interval and kept fixed across all replicates. We consider sample sizes $n=200, 500$.

We fit the data using the structured additive zero- and one-inflated quantile regression model proposed in Section \ref{sec:boundedquantreg} with the following structure
\begin{align*}
    \begin{split}
        \log\left( \dfrac{p_{0i}}{p_{2i}} \right) &=  \beta_{00} + f_{01}(x_{1i}) + f_{02}(x_{2i}) ,\\
        \log\left( \dfrac{p_{1i}}{p_{2i}} \right) &= \beta_{10} + f_{11}(x_{1i}) + f_{12}(x_{2i}),\\
        \text{logit}\left[ Q_{y_i|y_i \in (0,1)} ( \tau | \bm{x}_i) \right]  &= \beta_{20} + f_{21}(x_{1i}) + f_{22}(x_{2i}),\\ 
    \end{split} 
\end{align*}
for $i=1, \ldots, n$. We consider $\tau \in \{0.1, 0.5, 0.9\}$ and conduct $100$ replications. Posterior inference is carried out using the Liesel library in Python, as described in Section \ref{sec:computationalimplementation}. We use a warmup period of $1{,}500$ iterations, draw $5{,}000$ posterior samples, and run four parallel chains.

\begin{figure}[!ht]
    \centering
    \begin{minipage}{0.43\textwidth}
        \centering
        \includegraphics[width=\linewidth]{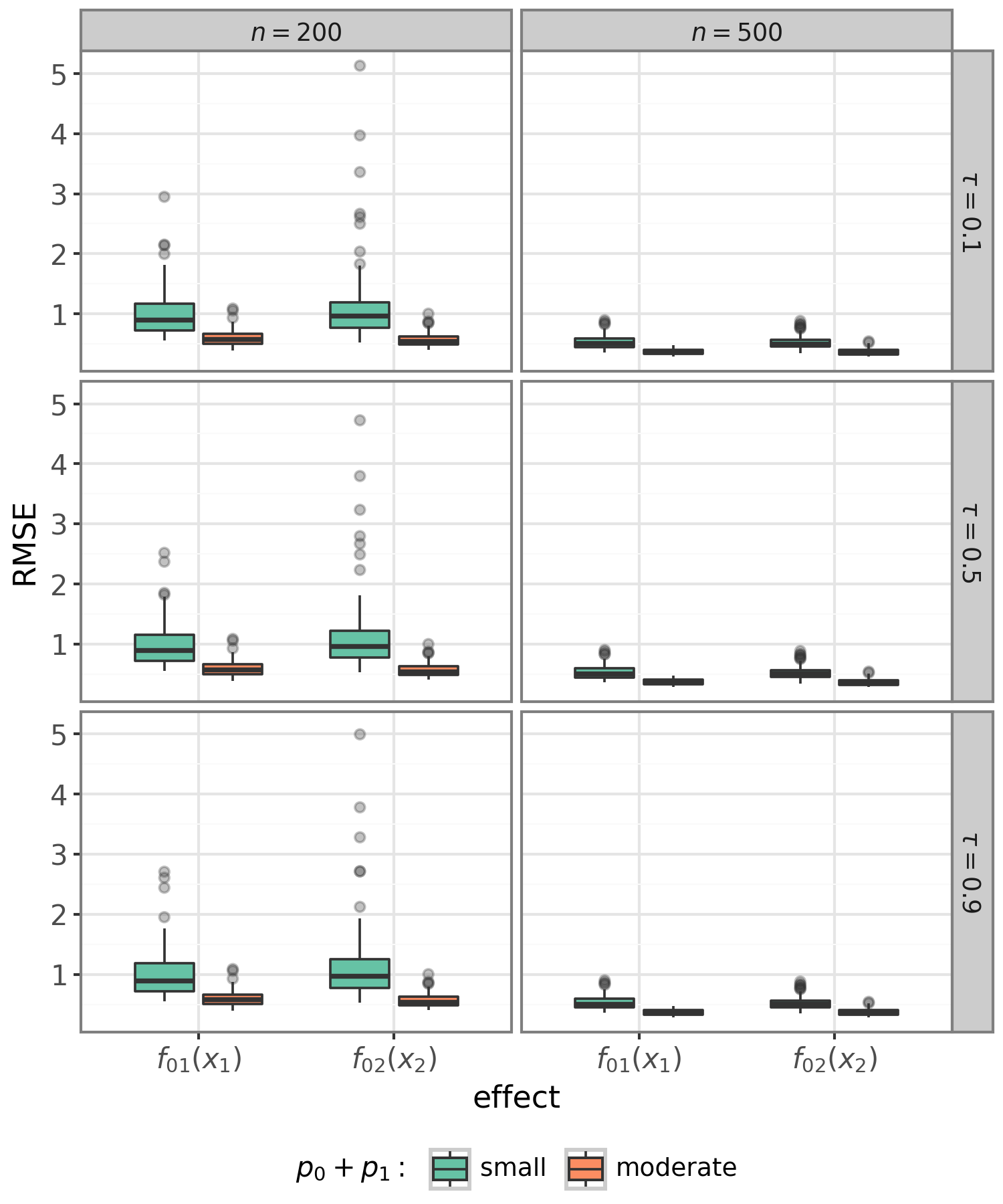}
    \end{minipage}\hfill
    \begin{minipage}{0.43\textwidth}
        \centering
        \includegraphics[width=\linewidth]{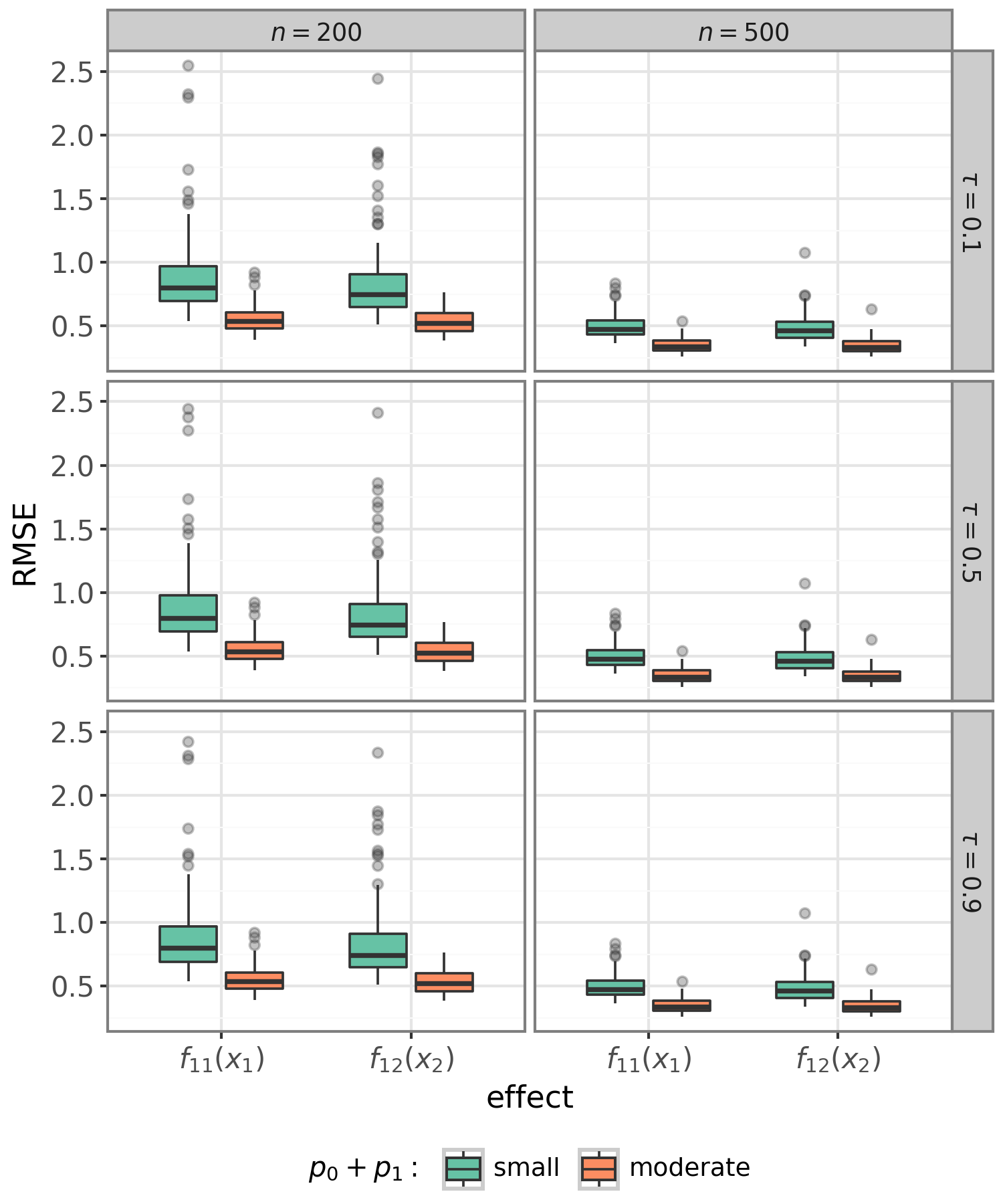}
    \end{minipage}\hfill
        \begin{minipage}{0.43\textwidth}
        \centering
        \includegraphics[width=\linewidth]{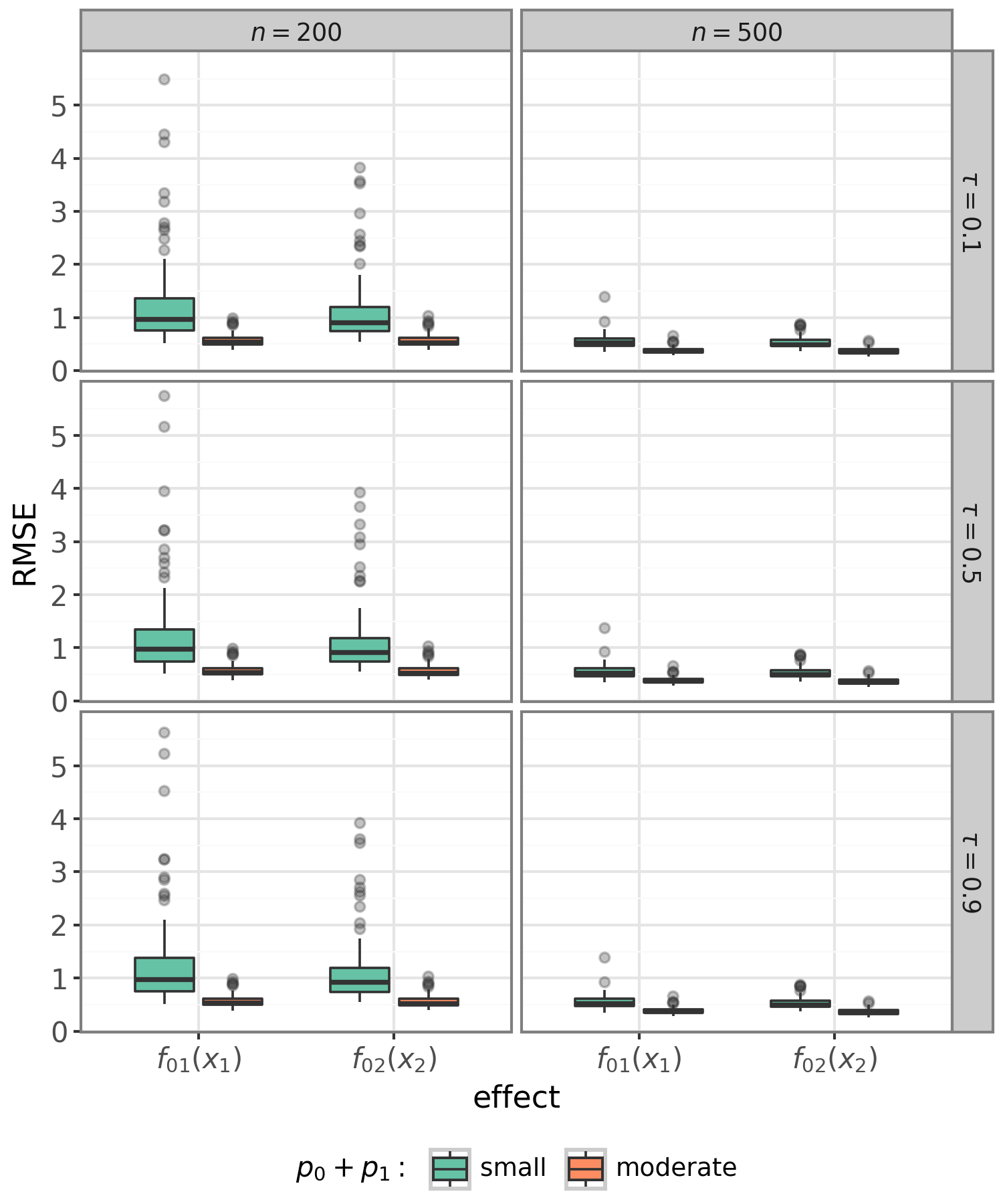}
    \end{minipage}\hfill
    \begin{minipage}{0.43\textwidth}
        \centering
        \includegraphics[width=\linewidth]{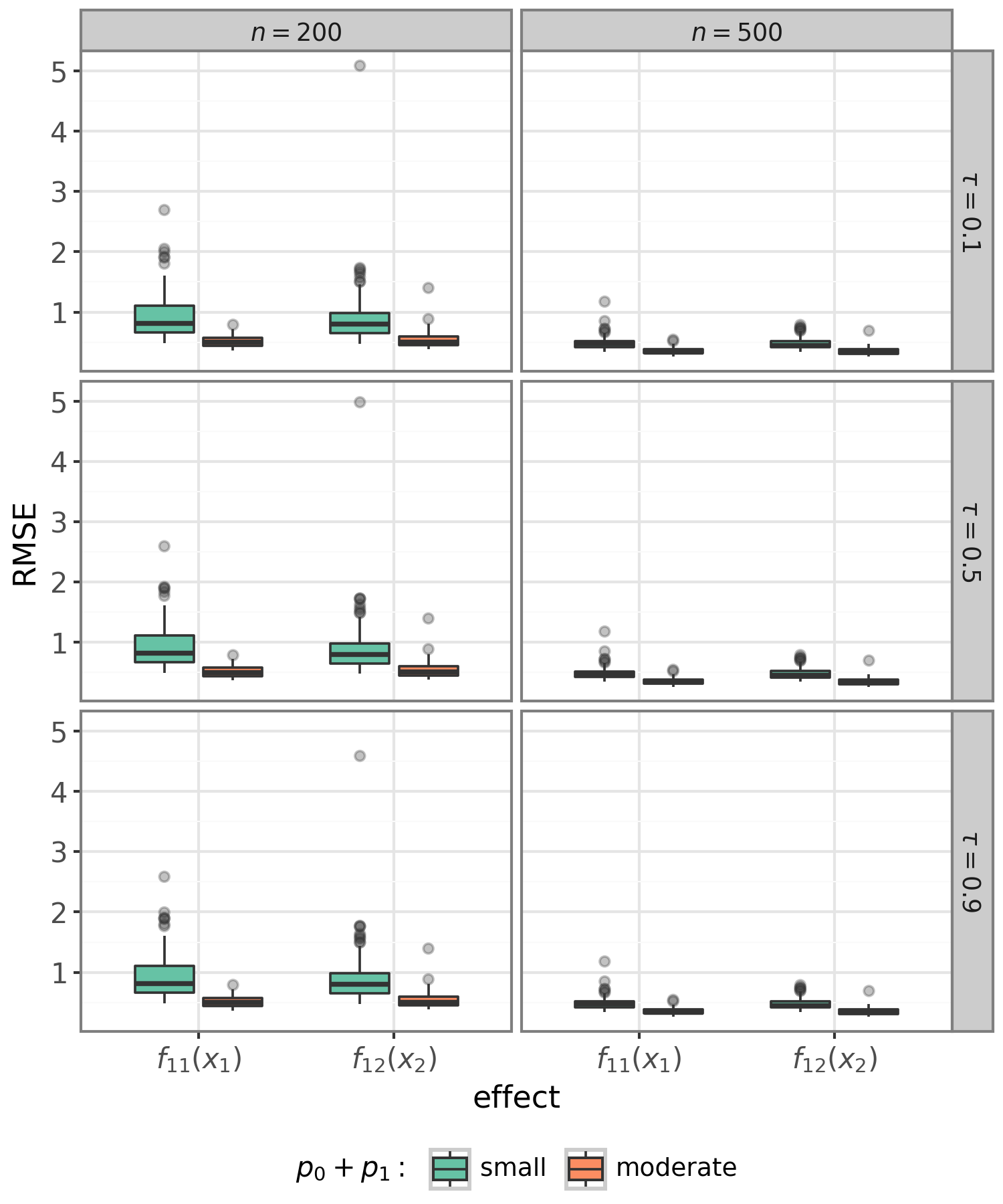}
    \end{minipage}\hfill
    \caption{Boxplot of the RMSE across all posterior samples for each covariate effect of the discrete part for Scenario S1 (top row) and Scenario S2 (bottom row).}
    \label{fig:discretepart}
\end{figure}

To evaluate the results, we compute the predicted function values and their confidence intervals on a 100-point equidistant grid spanning the range of the observed covariate values. 
For each posterior sample, we calculate the mean squared error (MSE) of the predicted functions as
\begin{align}\label{eq:MSE}
    \text{MSE}\left(\hat{f}^{[j]}(x_i)\right) = \dfrac{1}{r}\sum_{i=1}^r \left[f(x_i)- \hat{f}^{[j]}(x_i)\right]^2,
\end{align}
where $f$ and $\hat{f}^{[j]}(x_i)$ denote, respectively, the true and the predicted effect at the $i$th grid point for the $j$th posterior sample, and the index $i = 1, \dots, r$ identifies the grid points; $r=100$. Figure \ref{fig:discretepart} shows the plot of the root MSE (RMSE), averaged across all posterior samples for each covariate effect of the discrete component. 

The results show that, for a large sample size ($n=500$), the RMSEs of the effects are very close to zero. For a moderate sample size ($n=200$) and a small probability of observing $\{0,1\}$, some RMSEs are substantially larger. This is expected because, in this scenario, the response variable of the discrete component
\[
    (\mathbb{I}_{\{0\}}(y_i), \mathbb{I}_{\{1\}}(y_i), \mathbb{I}_{(0,1)}(y_i))
\]
is mostly $(0,0,1)$, providing very little information to estimate the discrete parameters. In contrast, when the probability of observing $\{0,1\}$ is moderate, the RMSEs remain close to zero even for $n=200$. Moreover, the RMSEs of the discrete component effects do not appear to be influenced by the choice of quantile being modeled, as expected.

\begin{figure}[!ht]
    \centering
    \begin{minipage}{0.45\textwidth}
        \centering
        \includegraphics[width=\linewidth]{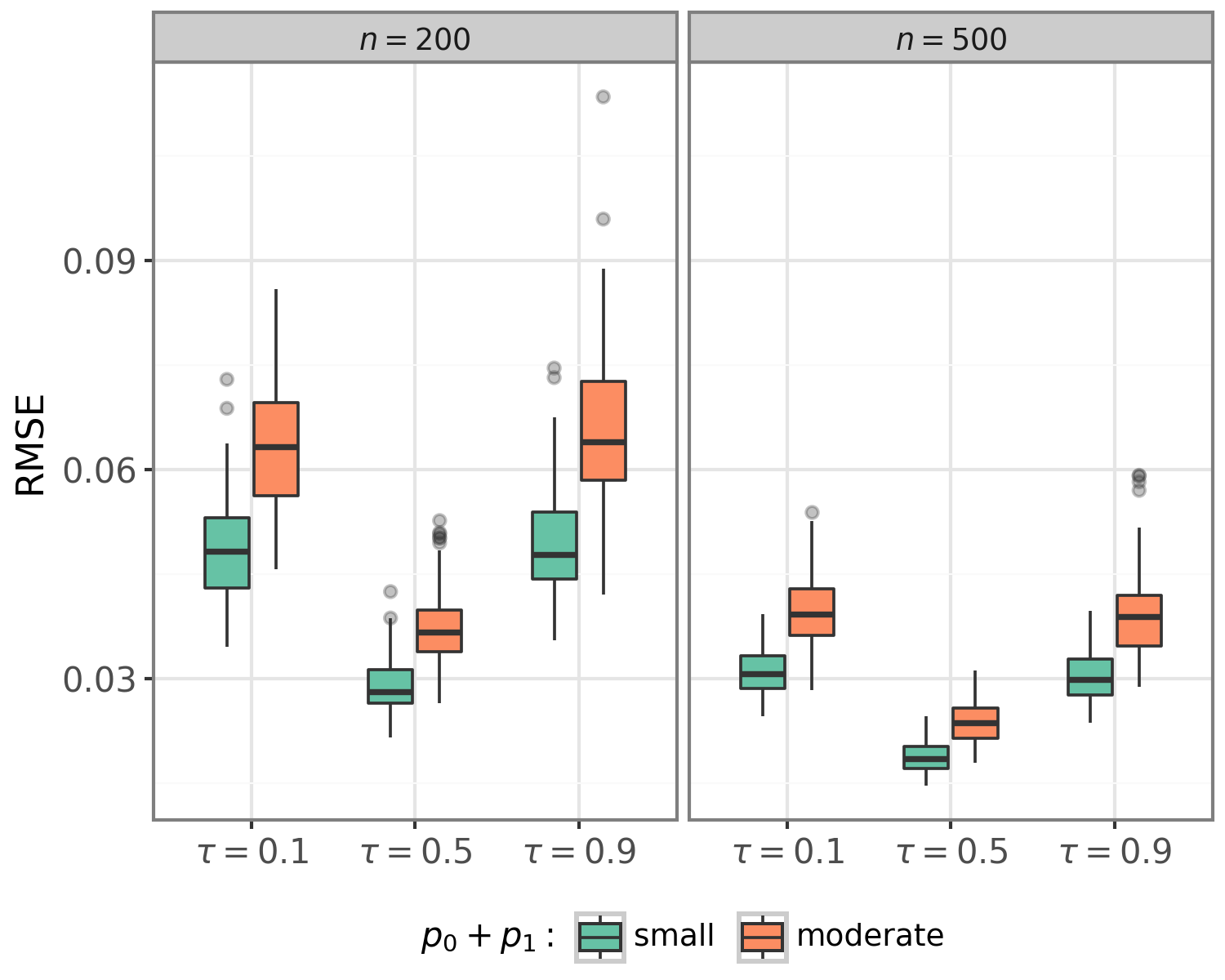}
    \end{minipage}\hfill
    \begin{minipage}{0.45\textwidth}
        \centering
        \includegraphics[width=\linewidth]{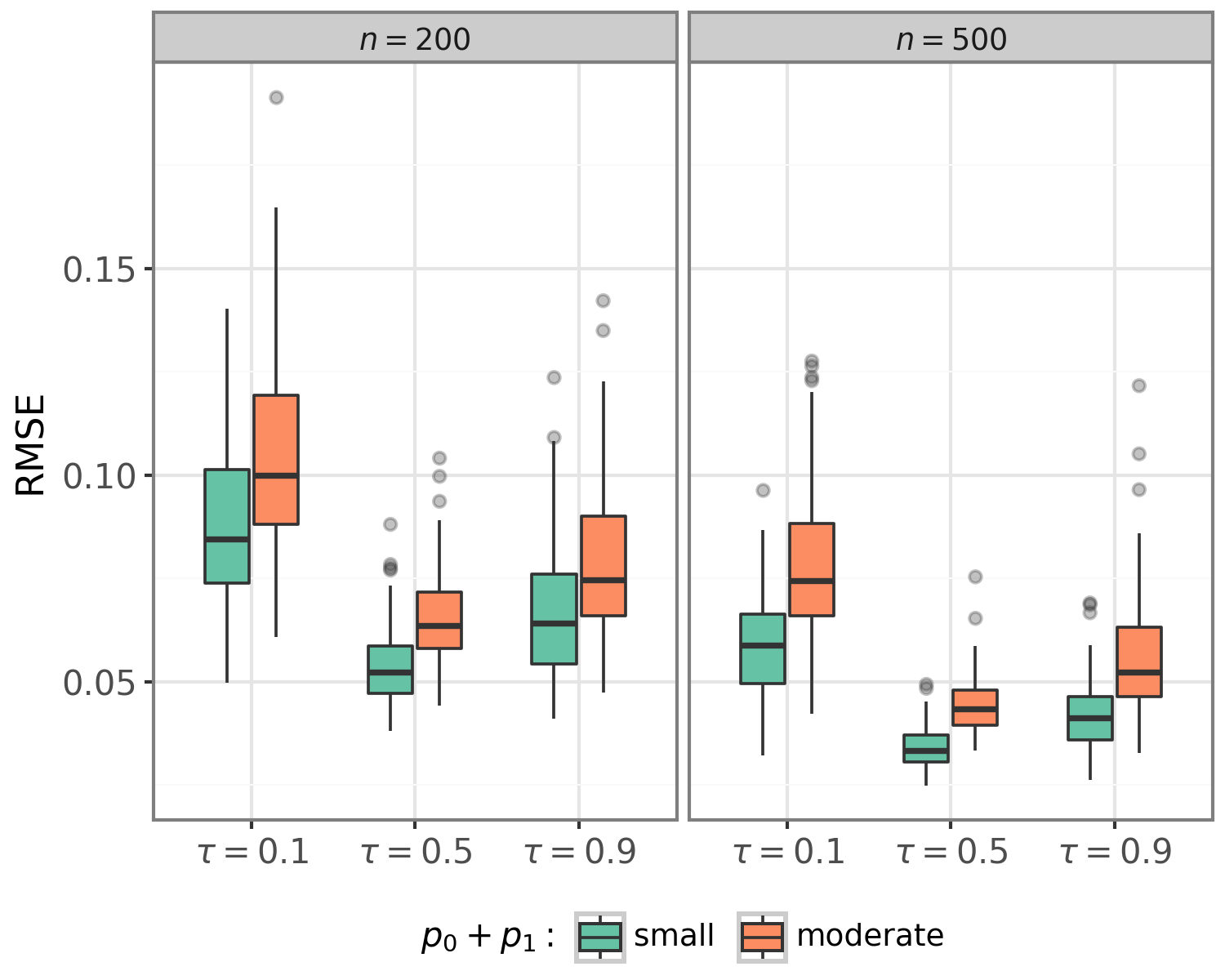}
    \end{minipage}
    \caption{Boxplot of the RMSE across all posterior samples for the predicted quantiles considering scenarios S1 (left) and S2 (right).}
    \label{fig:quantiles}
\end{figure}

To evaluate the estimation of the continuous component, we compute the RMSE across all posterior samples for the quantiles of the continuous outcome. The estimated $\tau$th conditional quantile of $y_i$ is given by 
$$\widehat{Q}_{y_i|y_i \in (0,1)} (\tau | \bm{x}_i) = \text{logit}^{-1}\left[  \hat{\beta}_{20} + \hat{f}_{21}(x_{1i}) + \hat{f}_{22}(x_{2i})\right].$$ 
The RMSE is then computed based on the predicted quantiles and the true quantiles defined in Equation~\eqref{eq:quantileGJSt}, using the new observations. Figure \ref{fig:quantiles} shows the RMSEs across all posterior samples for the predicted quantiles under both scenarios S1 and S2. In both scenarios, the RMSEs are smaller when modeling the median quantiles compared to the extreme quantiles. The RMSEs tend to be lower when the probability of observing $\{0,1\}$ is small, which is expected since in this case there are more observations in the continuous part. Finally, the RMSEs under Scenario S1 are smaller than under Scenario S2, which is also expected, as Scenario S1 corresponds to correctly specified models (the data are generated from the ALD). Nevertheless, the RMSEs under Scenario S2 remain low and close to zero, indicating that the model still provides a good fit even under likelihood misspecification.

\begin{figure}[!ht]
    \centering
    \begin{minipage}{0.8\textwidth}
        \centering
        \includegraphics[width=\linewidth]{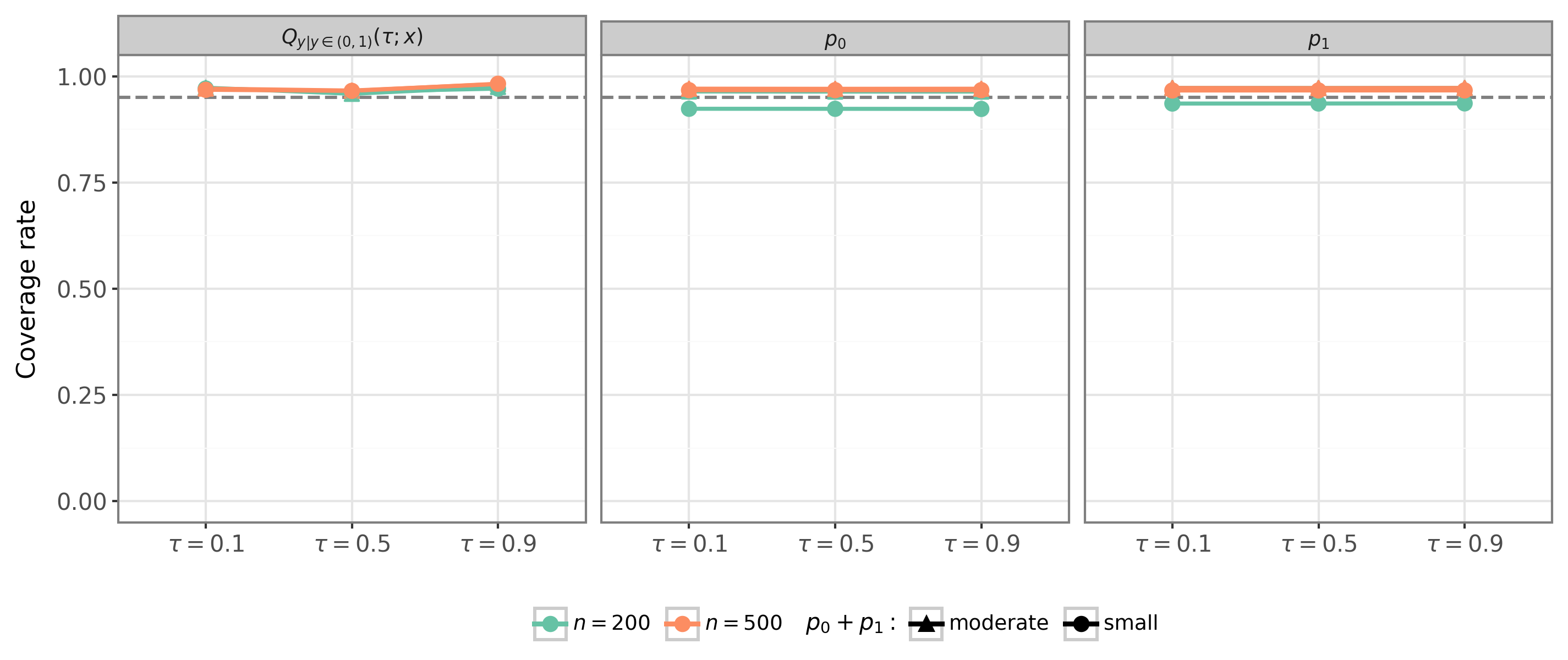}
    \end{minipage}\hfill
    \begin{minipage}{0.8\textwidth}
        \centering
        \includegraphics[width=\linewidth]{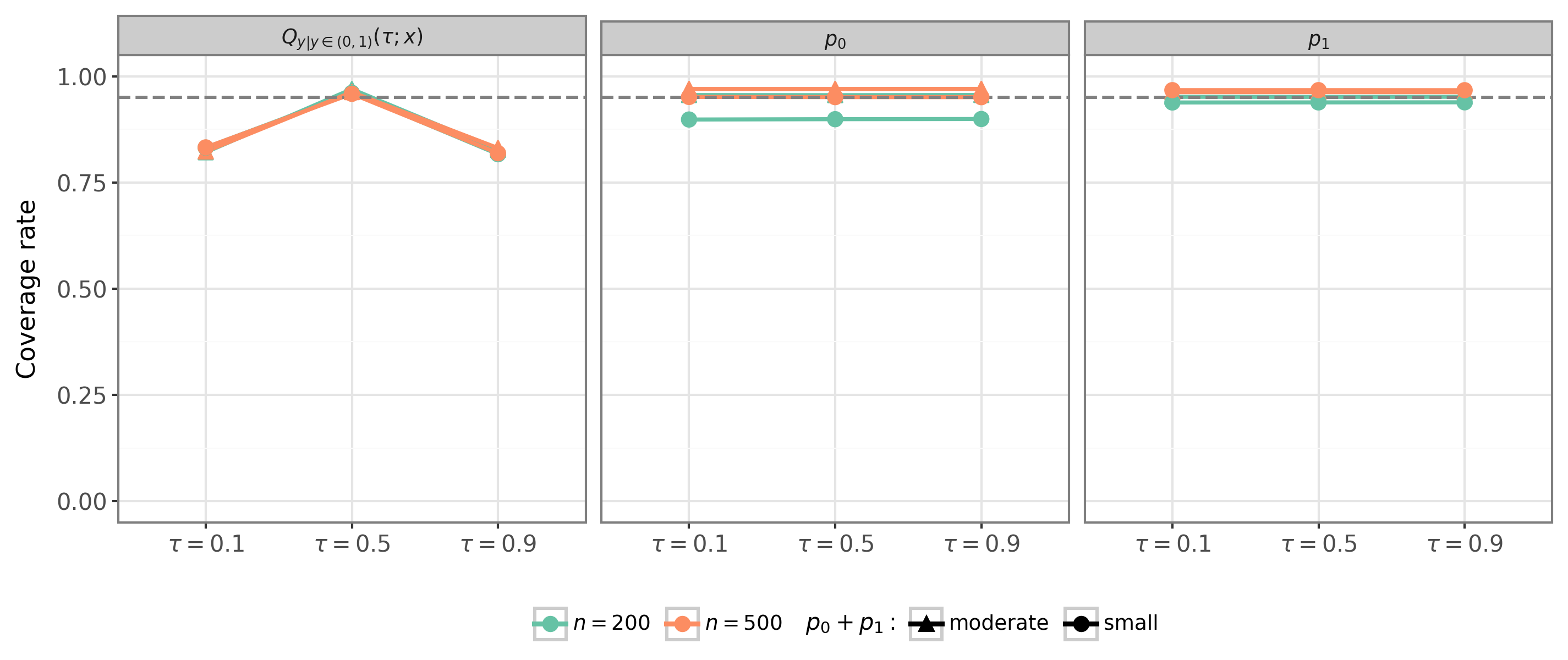}
    \end{minipage}\hfill
    \caption{Coverage rate of the $95\%$ confidence interval for the $\tau$th quantile, $p_{0i}$, and $p_{1i}$ for scenarios S1 (top row) and S2 (bottom row). The dashed gray line indicates the nominal $95\%$ coverage level.}
    \label{fig:cr}
\end{figure}

Finally, we compute the coverage rate of the $95\%$ credible interval for $Q_{y_i|y_i \in (0,1)} ( \tau | \bm{x}_i)$, $p_{0i}$, and $p_{1i}$ as follows. For each posterior sample, we compute the predicted values of these functions for the 100 test observations. For each replicate and each test observation, we calculate the 95\% credible intervals from the posterior samples and record whether the interval covers the true value ($1$ if yes, $0$ if no). We first average these indicators over all replicates to obtain a coverage rate for each test observation, and then average over all test observations to obtain an overall coverage rate. The results are presented in Figure \ref{fig:cr}.

Overall, for both scenarios S1 and S2, the coverage rates for $p_{0i}$ and $p_{1i}$ are close to the nominal level of $95\%$, with better performance observed for larger sample sizes and higher probabilities of observing $\{0,1\}$. The coverage rates for $Q_{y_i|y_i \in (0,1)} ( \tau | \bm{x}_i)$ are also close to the nominal level in scenario S1 for all quantiles, but they are noticeably below the nominal level when modeling extreme quantiles ($\tau = 0.1$ and $\tau = 0.9$) in scenario S2. Apart from this case, coverage rates are fairly consistent across different quantiles.

\section{Applications}\label{sec:applications}

We now present two empirical applications to illustrate the use of the proposed models. These examples are intended to demonstrate the flexibility of the modeling framework and its performance in different contexts. For both applications, we run four MCMC chains, each with $1{,}500$ warm-up iterations and $5{,}000$ posterior iterations. Posterior samples are collected without thinning, resulting in a total of $20{,}000$ samples for inference. Diagnostics indicate good mixing and no signs of convergence issues. The code for reproducing some of the results is available at the GitHub repository \href{https://github.com/ffqueiroz/inflated_quantreg}{inflated\_quantreg}.

\subsection{Traffic fatality data}\label{sec:application1}

We use data on mortality due to traffic accidents of $n = 5{,}499$ Brazilian municipal districts in the year 2002. The data were obtained from the DATASUS database (\url{https://datasus.saude.gov.br/}) and the IPEADATA database (\url{https://www.ipeadata.gov.br/}). The response variable ($y$) is the proportion of deaths caused by traffic accidents and the covariates are the logarithm of the number of inhabitants of the municipality (\texttt{lnpop}), the proportion of residents aged between 20 and 29 years (\texttt{prop2029}), the human development index of education of the municipal district (\texttt{hdie}), and the municipal district (\texttt{UF}). A subset of these data were first analyzed in \textcite{ospinaFerrari} using the zero-inflated beta regression models with linear predictors. In their analysis, the covariate \texttt{UF} was not included. The main objective of this study is to investigate the effect of the young population (\texttt{prop2029}) on the proportion of deaths caused by traffic accidents while accounting for potential confounding variables and the spatial structure captured by the covariate \texttt{UF}. 

\begin{figure}[!ht]
    \centering
    \begin{minipage}{0.9\textwidth}
        \centering
        \includegraphics[width=\linewidth]{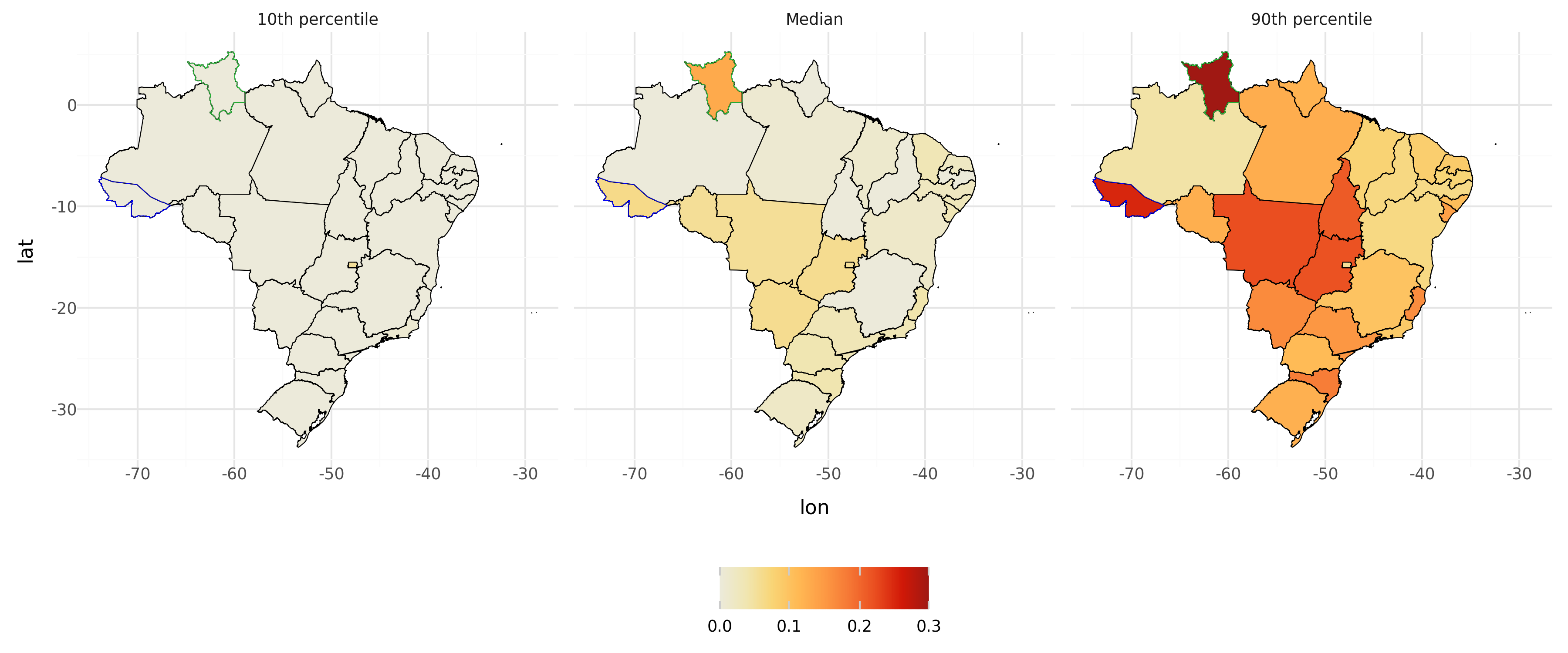}
    \end{minipage}\hfill
    \caption{Spatial distribution of the sample quantiles of $y$ across Brazilian states (UF): 10th percentile ($\tau = 0.1$, left panel), median ($\tau = 0.5$, center panel), and 90th percentile ($\tau = 0.9$, right panel).}
    \label{fig:sample-quantiles}
\end{figure}

The response variable exhibits a markedly right-skewed distribution, with approximately $38\%$ of municipal districts reporting zero deaths due to traffic accidents. While the median proportion is $0.024$, the mean reaches $0.048$, indicating the influence of municipalities with relatively high mortality rates. The maximum observed value is $0.80$, occurring in a municipal district in the state of São Paulo. Figure \ref{fig:sample-quantiles} presents the sample quantiles of $y$ by the municipal districts. Figure \ref{fig:sample-quantiles} shows that the 10th percentile of $y$ is close to zero for all states, indicating that many municipalities report very low or no deaths due to traffic accidents. The median of $y$ is around $0.05$ for most states, with the state Roraima, whose border is highlighted in green, showing a slightly higher median of approximately $0.13$. The 90th percentile reaches values near $0.3$ in some states, including Roraima, Acre (highlighted in blue), and several states in the Central-West region, indicating that certain municipalities in these regions experience higher mortality rates.

We model these data using the zero-inflated structured additive quantile regression with the following specifications
\begin{align} \label{eq:predictors-ap1}
    \begin{split}
        \text{logit}(p_i) &= \beta_{10} 
            + f_{11}(\texttt{lnpop})
            + f_{12}(\texttt{prop2029})
            + f_{13}(\texttt{hdie})
            + f_{1,\text{spatial}}(\texttt{UF}),\\
        \text{logit}\left[Q_{y_i\mid y_i \in (0,1)}(\tau \mid \mathbf{x}_i)\right] &= \beta_{20}
            + f_{21}(\texttt{lnpop})
            + f_{22}(\texttt{prop2029})
            + f_{23}(\texttt{hdie})
            + f_{2,\text{spatial}}(\texttt{UF}),
    \end{split}
\end{align}
where $p_i = \mathbb{P}(y_i=0)$ is the probability of observing zero, $Q_{y_i\mid y_i \in (0,1)}(\tau \mid \mathbf{x}_i)$ is the $\tau$th quantile of the conditional distribution of $y_i$ given that $y_i \in (0,1)$, $\beta_{10}$ and $\beta_{20}$ are the intercepts and $f_{j1}$, $f_{j2}$, and $f_{j3}$, $j=1,2$, are the nonlinear effects of the covariates \texttt{lnpop}, \texttt{prop2029}, \texttt{hdie}, respectively, which are modeled using cubic penalized splines with $20$ basis functions and a weak inverse-Gamma prior controlling smoothness (shape and scale equal to $0.01$). The spatial effects $f_{j,\text{spatial}}(\texttt{UF})$, $j=1,2$, are modeled using a Gaussian Markov random field defined over the $27$ Brazilian states. The neighborhood structure is defined based on the spatial adjacency of the regions, so that adjacent states are treated as neighbors. Smoothness of the spatial effects is controlled by a variance parameter with a weak inverse-gamma prior with both shape and scale equal to $0.01$.

For the continuous part, we model the $5\%$,  $10\%$, $20\%$, $30\%$, $50\%$, $70\%$, $80\%$, $90\%$, and $95\%$ conditional quantiles. This choice enables us to assess not only the central pattern observed across municipalities, provided by the median, but also the behavior of municipalities with the smallest and largest proportions of deaths due to traffic accidents, among those reporting at least one traffic fatality. Additionally, the zero-inflated structure explicitly models the probability of observing zero deaths, allowing the model to account for municipalities with no traffic fatalities.

\begin{figure}[!ht]
    \centering
    \begin{minipage}{0.33\textwidth}
        \centering
        \includegraphics[width=\linewidth]{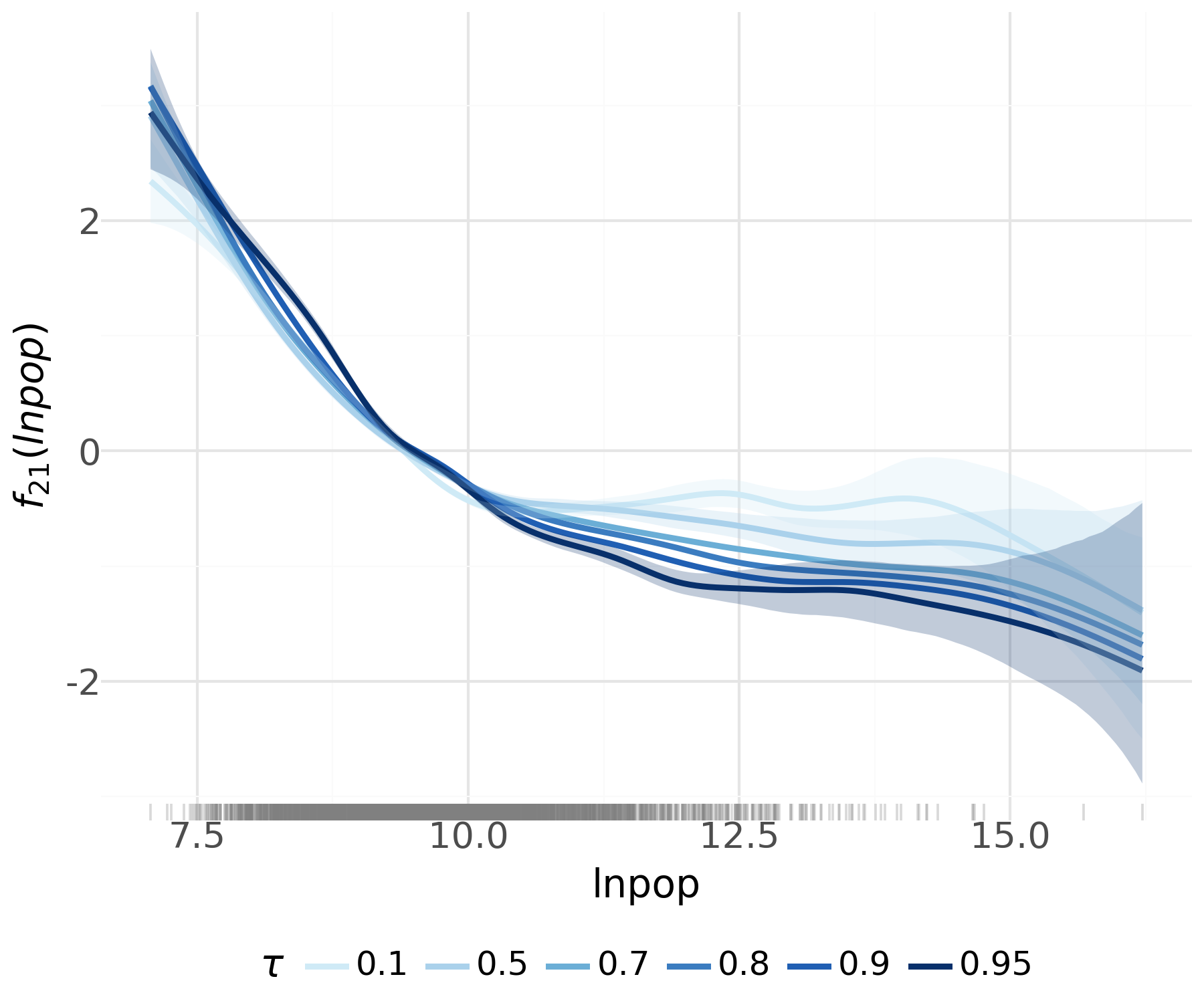}
    \end{minipage}\hfill
    \begin{minipage}{0.33\textwidth}
        \centering
        \includegraphics[width=\linewidth]{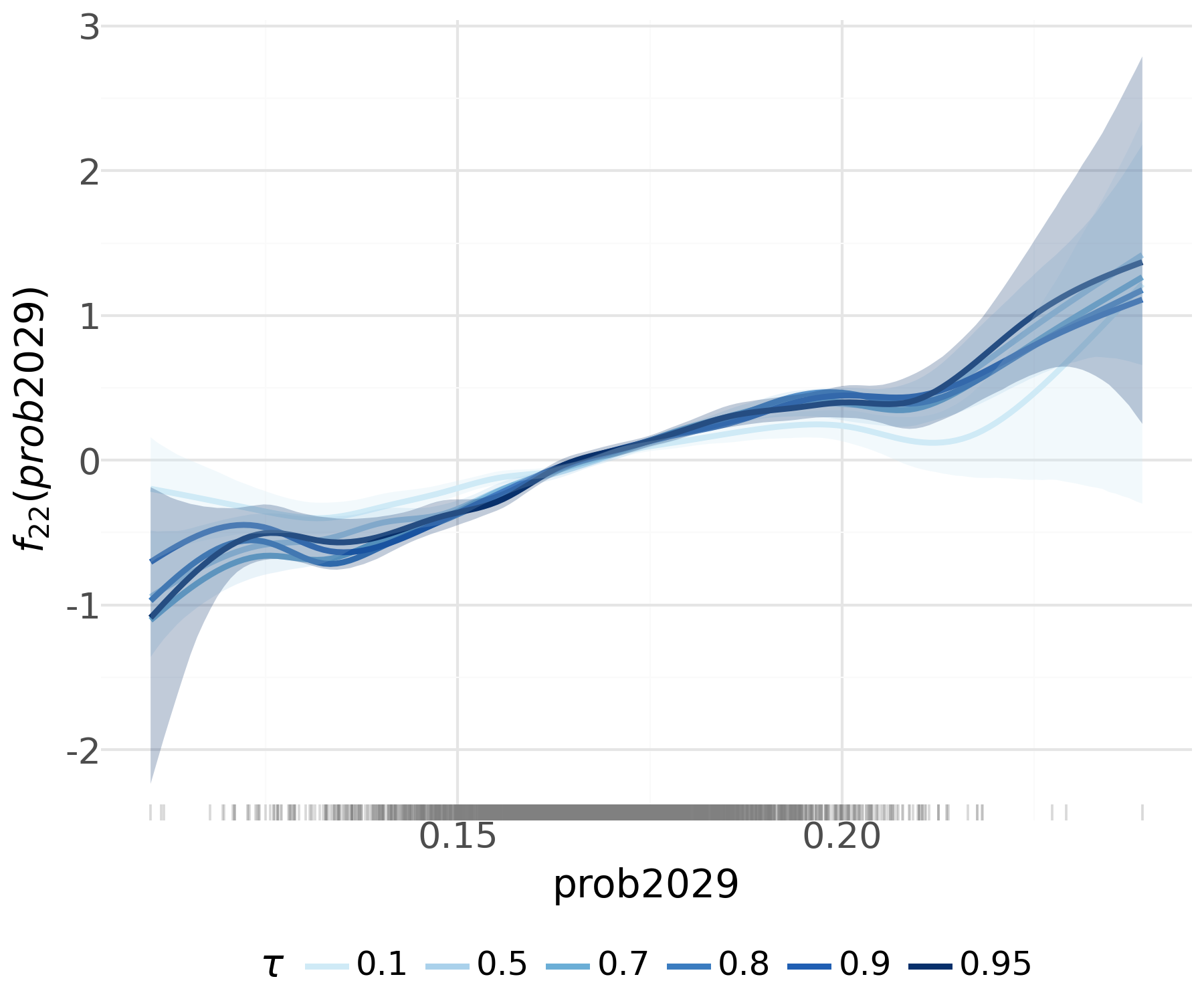}
    \end{minipage}\hfill
    \begin{minipage}{0.33\textwidth}
        \centering
        \includegraphics[width=\linewidth]{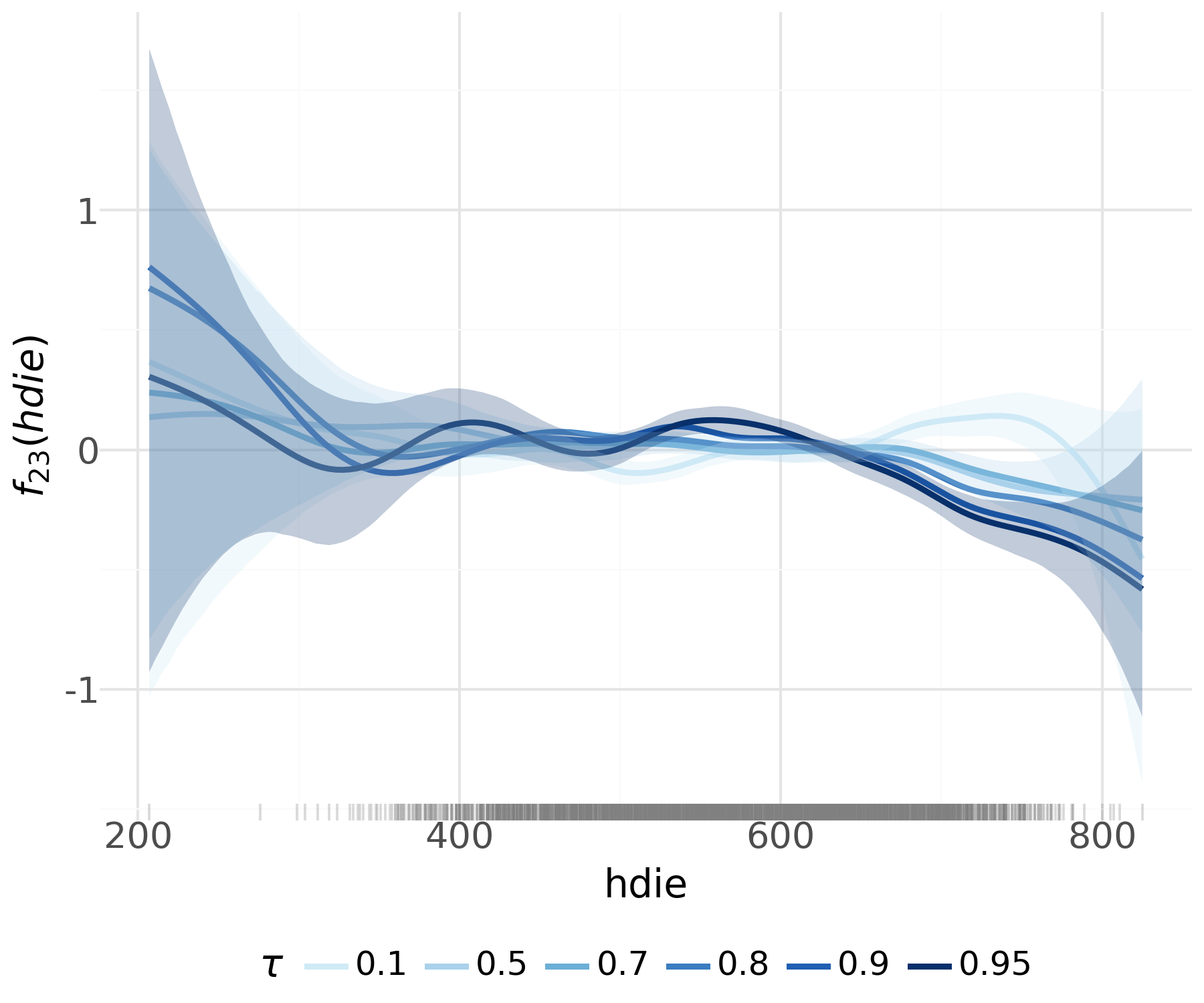}
    \end{minipage}\hfill
    \begin{minipage}{\textwidth}
        \centering
        \includegraphics[width=\linewidth]{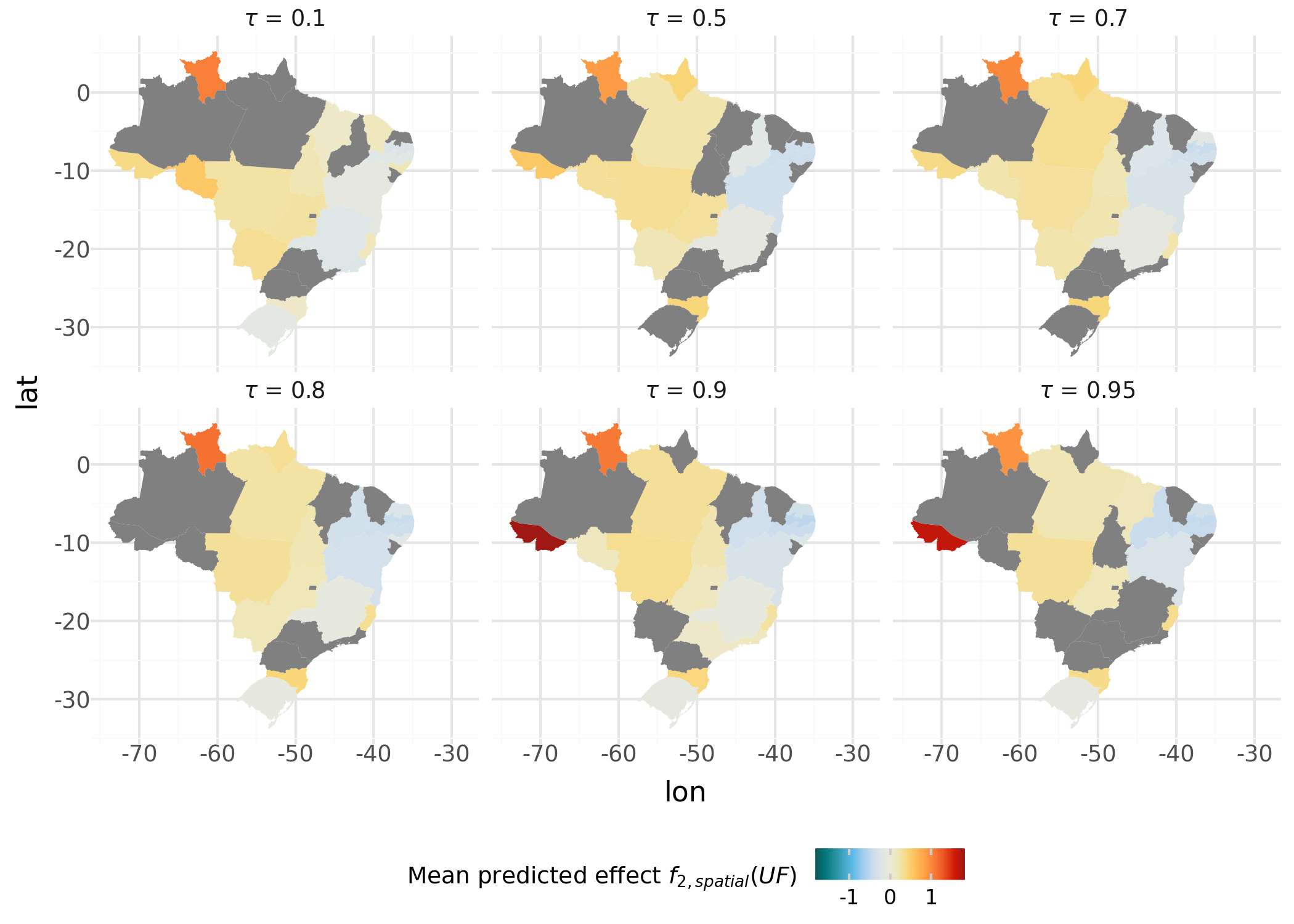}
    \end{minipage}\hfill
    \caption{Plots of the estimated effects of the continuous part. The top row shows the estimated nonlinear effects along with the $90\%$ credible interval. For visualization purposes, the credible interval are only plotted for the $10\%$, $50\%$, and $90\%$ quantiles. The middle and bottom rows present the estimated spatial effects for different quantiles. States whose $90\%$ credible intervals for the spatial effects include zero (nonsignificant) are shown in gray.}
    \label{fig:effects-cont}
\end{figure}

Figure \ref{fig:effects-cont} presents the plots of the estimated nonlinear effects of the continuous part as well as the estimated spatial effects. 
Here, we present results for the six selected quantiles; 
results for the remaining quantiles are in Section 2 of the Supplementary Material, as they closely resemble those for the 10\% quantile. Recall that, since the continuous component models only observations on $(0,1)$, these effects should be understood conditionally on the response variable taking values in this interval, i.e., excluding municipalities with zero deaths. That is, the modeled quantiles refer only to municipalities where at least one death due to traffic accidents was observed.

The estimated effect for the \texttt{lnpop} suggests a generally negative association between municipality size and the proportion of deaths due to traffic accidents across all conditional quantiles. Overall, municipalities with larger populations tend to exhibit a lower share of traffic-related mortality. The decline is most evident at smaller population sizes, indicating that the transition from very small to moderately sized municipalities is associated with relatively larger reductions in traffic mortality. At higher population levels, the effect becomes less pronounced, suggesting diminishing changes as population size increases. Although the direction of the association is consistent across quantiles, its magnitude varies. In particular, the negative effect of population size appears more pronounced at the upper quantiles ($80\%$, $90\%$, and $95\%$) than at the median or lower quantiles, suggesting that population size has an association with reducing traffic mortality in municipalities with relatively higher proportions of traffic-related deaths. In practical terms, larger municipalities appear to better mitigate high levels of traffic fatalities, potentially reflecting differences in infrastructure, traffic enforcement, and access to healthcare.

A positive association is observed between the proportion of residents aged 20–29 (\texttt{prop2029}) and the proportion of traffic-related deaths across all quantiles. Municipalities with a larger share of young residents tend to experience higher traffic mortality when fatalities occur. The relationship is approximately linear but can be roughly divided into three segments: the first and third segments show a slightly faster rate of increase, while the middle segment exhibits a somewhat slower rate of increase, all within an overall upward trend. This pattern suggests that in municipalities with a relatively low share of young residents, even small increases in the proportion of young people are associated with noticeable increases in traffic mortality. In municipalities with a moderate share of young residents, additional increases have a smaller impact, whereas in municipalities with a relatively high share, further increases again lead to perceptible growth in traffic fatalities.

The estimated effect of \texttt{hdie} is very close to zero for all quantiles over a wide range of values, approximately from $200$ to $630$, indicating no clear association between the educational development index and the proportion of traffic-related deaths in municipalities with low to moderate levels of educational development. Beyond this range, the effect becomes more heterogeneous across quantiles. For the lower quantile ($\tau = 0.1$), the effect shows a slight increase at higher values of \texttt{hdie}, followed by a decline at the highest levels. In contrast, for the median and upper quantiles ($\tau \geq 0.5$), the effect becomes negative after \texttt{hdie} exceeds approximately 630, with the decline being more pronounced for higher quantiles, particularly at $\tau = 0.9$ and $\tau = 0.95$. This pattern suggests that higher levels of educational development are associated with lower proportions of traffic-related deaths among municipalities with the largest values of traffic mortality.

The estimated spatial effects reveal regional patterns in traffic mortality across Brazil. Acre and Roraima, located in the Northern region, show consistently positive spatial effects, particularly at higher quantiles, indicating that municipalities in these states tend to have relatively higher proportions of traffic-related deaths, conditional on at least one fatality occurring. In contrast, S\~ao Paulo, the country's most populous and highly urbanized state, exhibits effects close to zero across all quantiles. Beyond these specific cases, spatial effects vary across regions: the Northern region generally shows positive effects, the Northeast tends to exhibit negative or heterogeneous effects, particularly at higher quantiles, the Central-West shows mostly neutral effects, and the Southern region displays small negative or near-zero effects. Overall, the magnitude of spatial effects is more pronounced at higher quantiles, indicating that regional heterogeneity is stronger in municipalities with relatively higher traffic mortality.

\begin{figure}[!ht]
    \centering
    \begin{minipage}{0.4\textwidth}
        \centering
        \includegraphics[width=\linewidth]{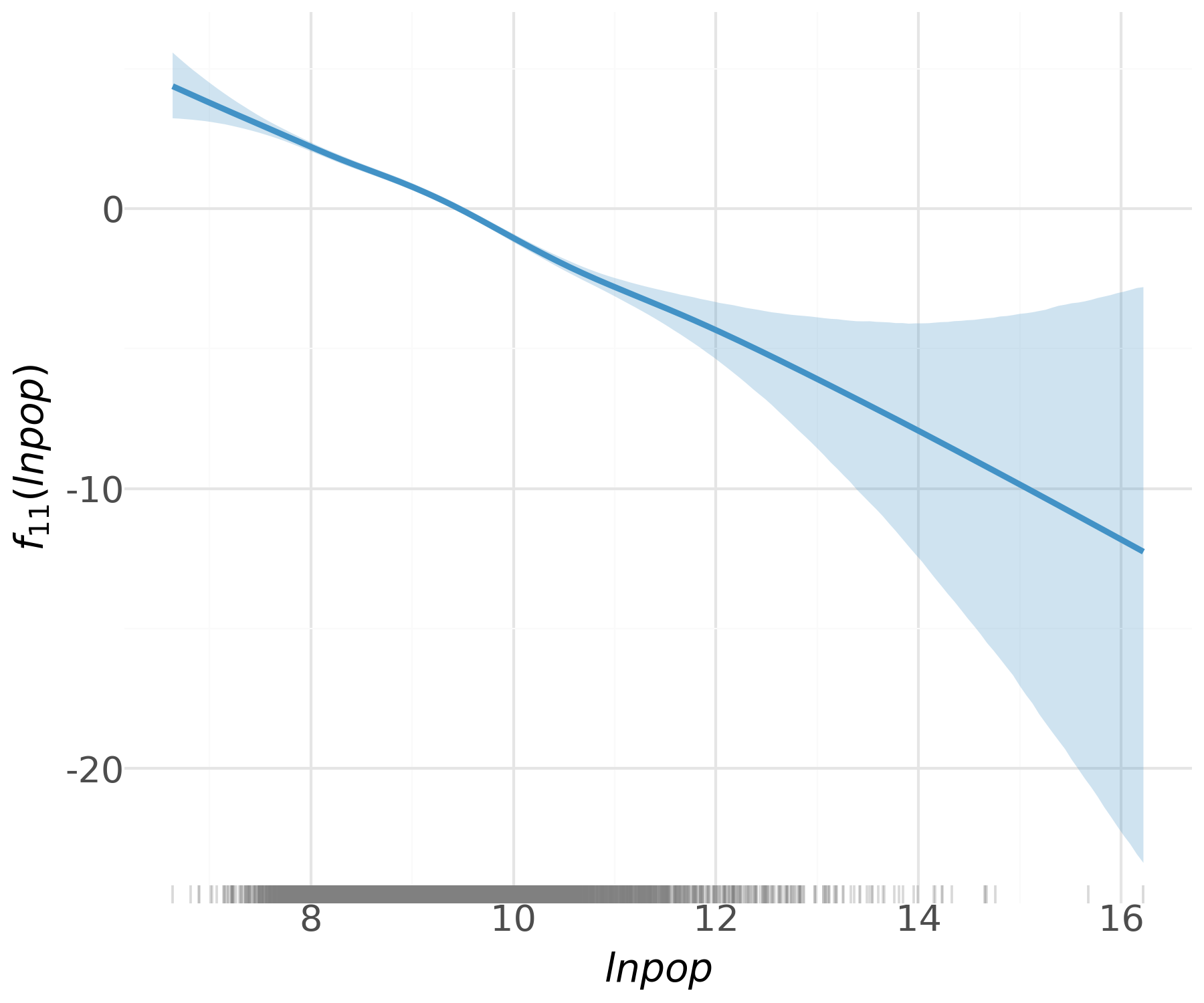}
    \end{minipage}
    \begin{minipage}{0.4\textwidth}
        \centering
        \includegraphics[width=\linewidth]{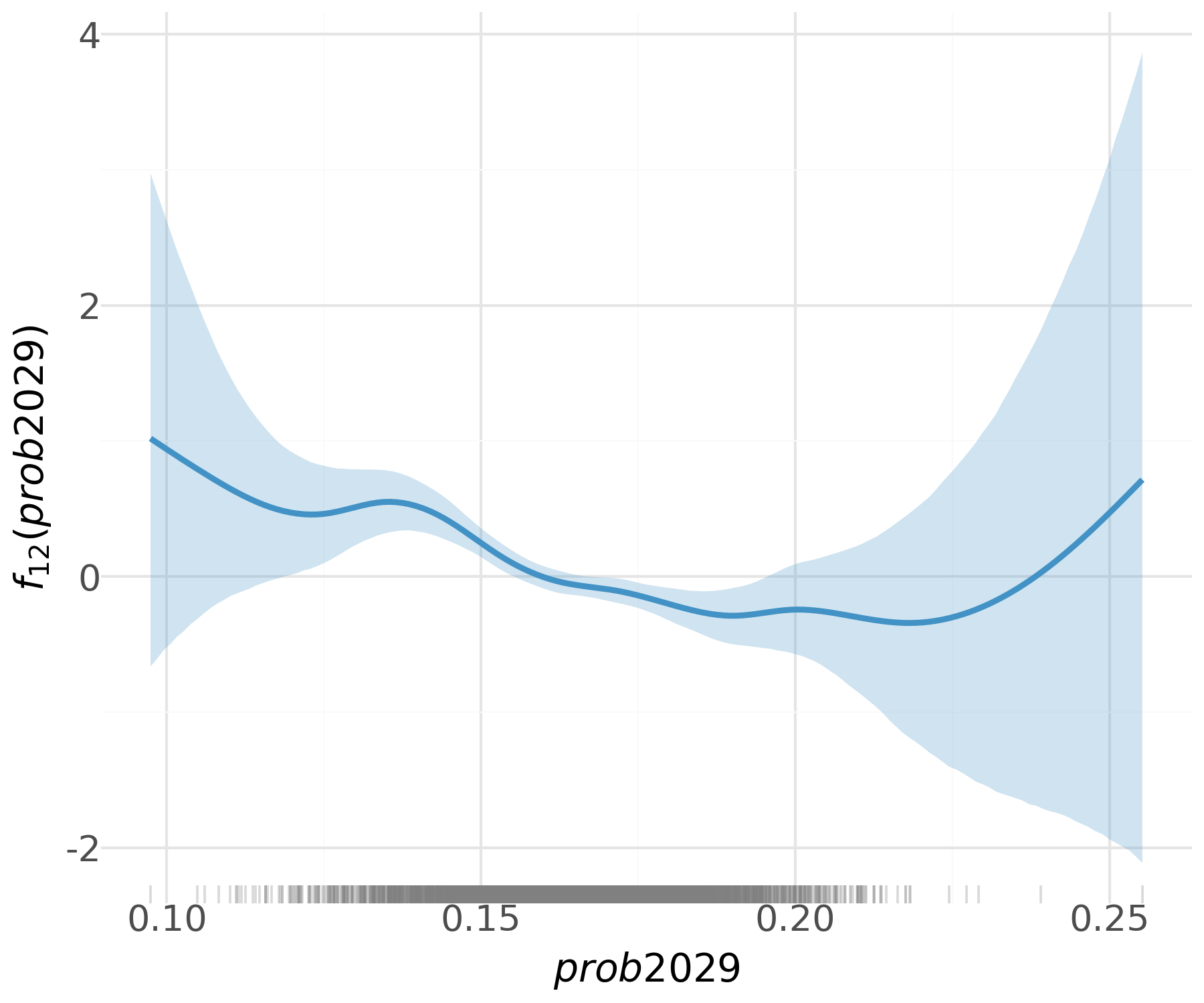}
    \end{minipage}
    \begin{minipage}{0.4\textwidth}
        \centering
        \includegraphics[width=\linewidth]{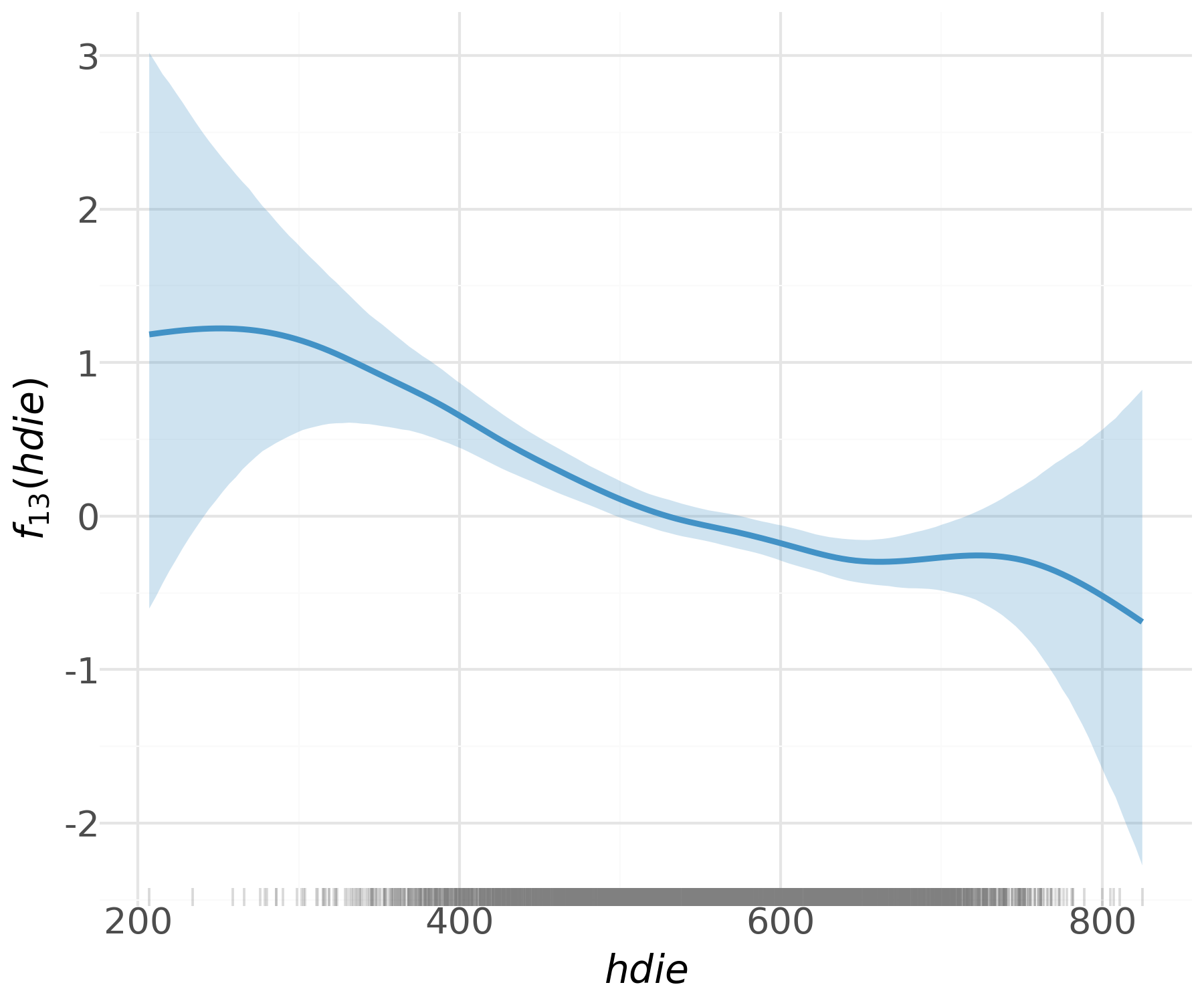}
    \end{minipage}
    \begin{minipage}{0.4\textwidth}
        \centering
        \includegraphics[width=\linewidth]{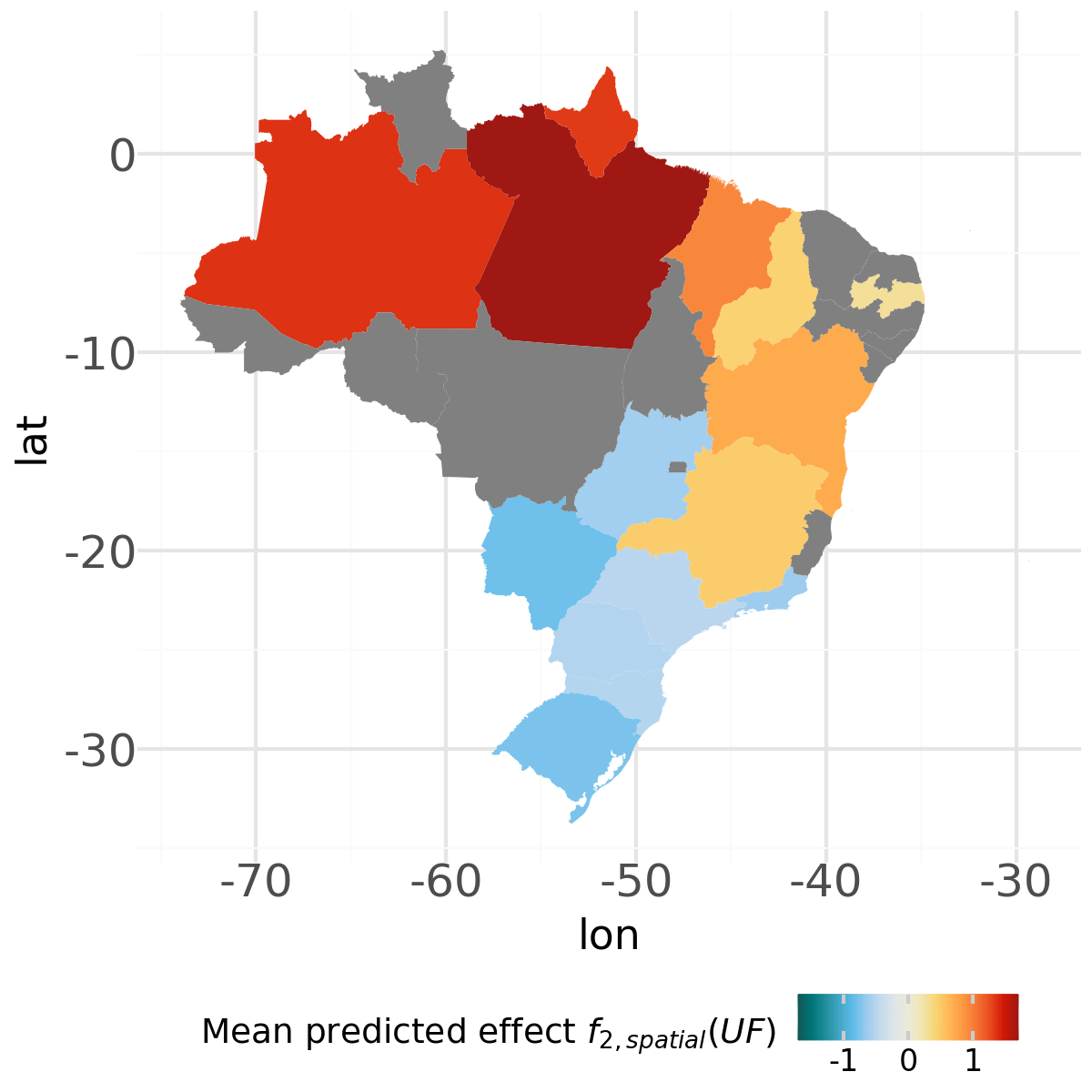}
    \end{minipage}
    \caption{Plots of the estimated effects of the discrete part. The bottom-right panel shows the estimated spatial effect, while the estimated nonlinear effects are shown in the other panels. States whose 90\% credible intervals for the spatial effects include zero (nonsignificant) are shown in gray.}
    \label{fig:effects-disc}
\end{figure}

Figure \ref{fig:effects-disc} presents the estimated effects of the discrete part. The estimated effect of \texttt{lnpop} is decreasing and approximately linear. This suggests that municipalities with higher populations are less likely to report zero deaths due to traffic accidents. The estimated effect of \texttt{prop2029} is close to zero across most values, with a small positive peak between $0.13$ and $0.15$, suggesting that municipalities with a young population share in this range may have a slightly higher probability of reporting zero traffic fatalities. A decreasing trend is observed in the estimated effect of \texttt{hdie}. For municipalities with low educational development (\texttt{hdie} below $500$), the effect is slightly positive, suggesting a higher probability of reporting zero deaths. For municipalities with moderate to high \texttt{hdie} (above $500$), the effect turns slightly negative but remains very close to zero, indicating minimal influence. For the spatial effects on the probability of zero traffic fatalities, positive effects are observed in most states of the Northern and Northeastern regions, suggesting that municipalities in these areas tend to have a higher chance of reporting zero deaths. Negative effects are seen in the South, Southeast, and parts of the Central-West, pointing to a lower probability of zero fatalities. These results highlight that municipalities in certain Northern and Northeastern states tend to experience fewer traffic fatalities, while more urbanized and densely populated areas are more likely to record at least one death.

Overall, when combining the results for continuous and zero-inflated parts, the results suggests that although traffic fatalities tend to be more common in larger municipalities, they represent a smaller share of overall mortality. The proportion of young residents has little influence on the probability of zero fatalities but is positively associated with traffic mortality in the continuous component, suggesting that municipalities with younger populations tend to experience a larger share of traffic-related deaths when fatalities occur. The educational development index shows only a weak association with the probability of zero fatalities but is negatively associated with the proportion of traffic-related deaths at higher quantiles, suggesting that higher educational development is associated with smaller nonzero share of  traffic-related deaths. Finally, the spatial effects reveal important regional disparities. Municipalities in Northern and Northeastern states tend to have higher probabilities of reporting zero fatalities, but when fatalities do occur, some of these states, particularly in the North, exhibit higher conditional mortality levels. Conversely, more urbanized regions such as the South and Southeast tend to have a lower probability of zero fatalities but also generally lower nonzero traffic-mortality levels.

For the sake of comparison, we also fit the zero-inflated beta regression model \parencite{ospinaFerrari} to this data. We consider the following specifications for the predictors of the conditional mean ($\mu_i$), conditional precision ($\phi_i$), and probability of zero ($p_i$)
\begin{align*}
    \begin{split}
        \text{logit}(p_i) &= \beta_{10} 
            + f_{11}(\texttt{lnpop})
            + f_{12}(\texttt{prop2029})
            + f_{13}(\texttt{hdie})
            + f_{1,\text{spatial}}(\texttt{UF}),\\
        \text{logit}(\mu_i) &= \beta_{20,\mu}
            + f_{21,\mu}(\texttt{lnpop})
            + f_{22,\mu}(\texttt{prop2029})
            + f_{23,\mu}(\texttt{hdie})
            + f_{2,\mu,\text{spatial}}(\texttt{UF}),\\
        \text{log}(\phi_i) &= \beta_{20,\phi}
            + f_{21,\phi}(\texttt{lnpop})
            + f_{22,\phi}(\texttt{prop2029})
            + f_{23,\phi}(\texttt{hdie})
            + f_{2,\phi,\text{spatial}}(\texttt{UF}).
    \end{split}
\end{align*}
We fit this model using Liesel, retaining the same functional forms, number of basis functions, spatial effects, and prior specifications for the regression parameters as in Equation~\eqref{eq:predictors-ap1} for the zero-inflated structured additive quantile regression model. 

\begin{figure}[!ht]
    \centering
    \begin{minipage}{0.32\textwidth}
        \centering
        \includegraphics[width=\linewidth]{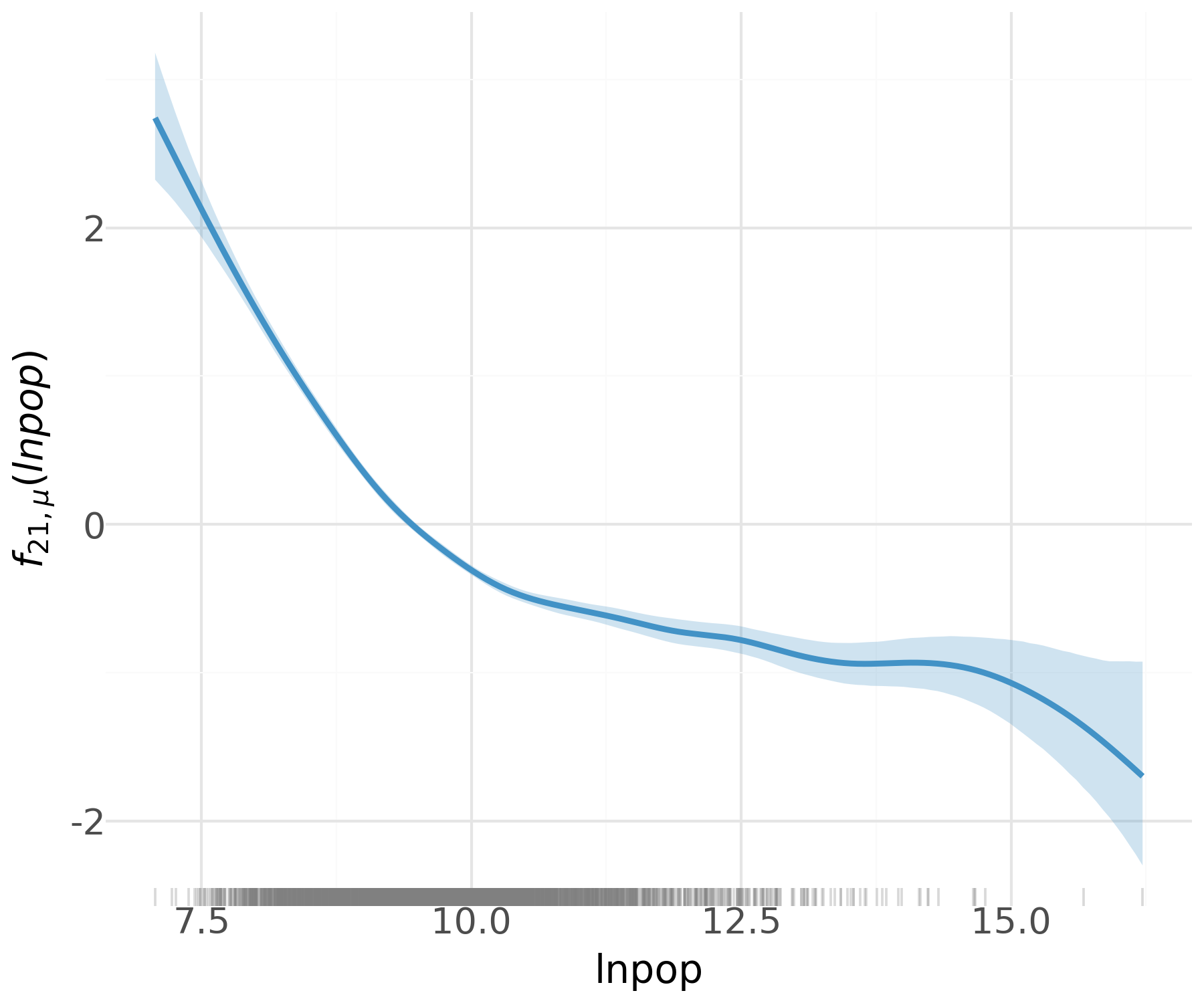}
    \end{minipage}
    \begin{minipage}{0.32\textwidth}
        \centering
        \includegraphics[width=\linewidth]{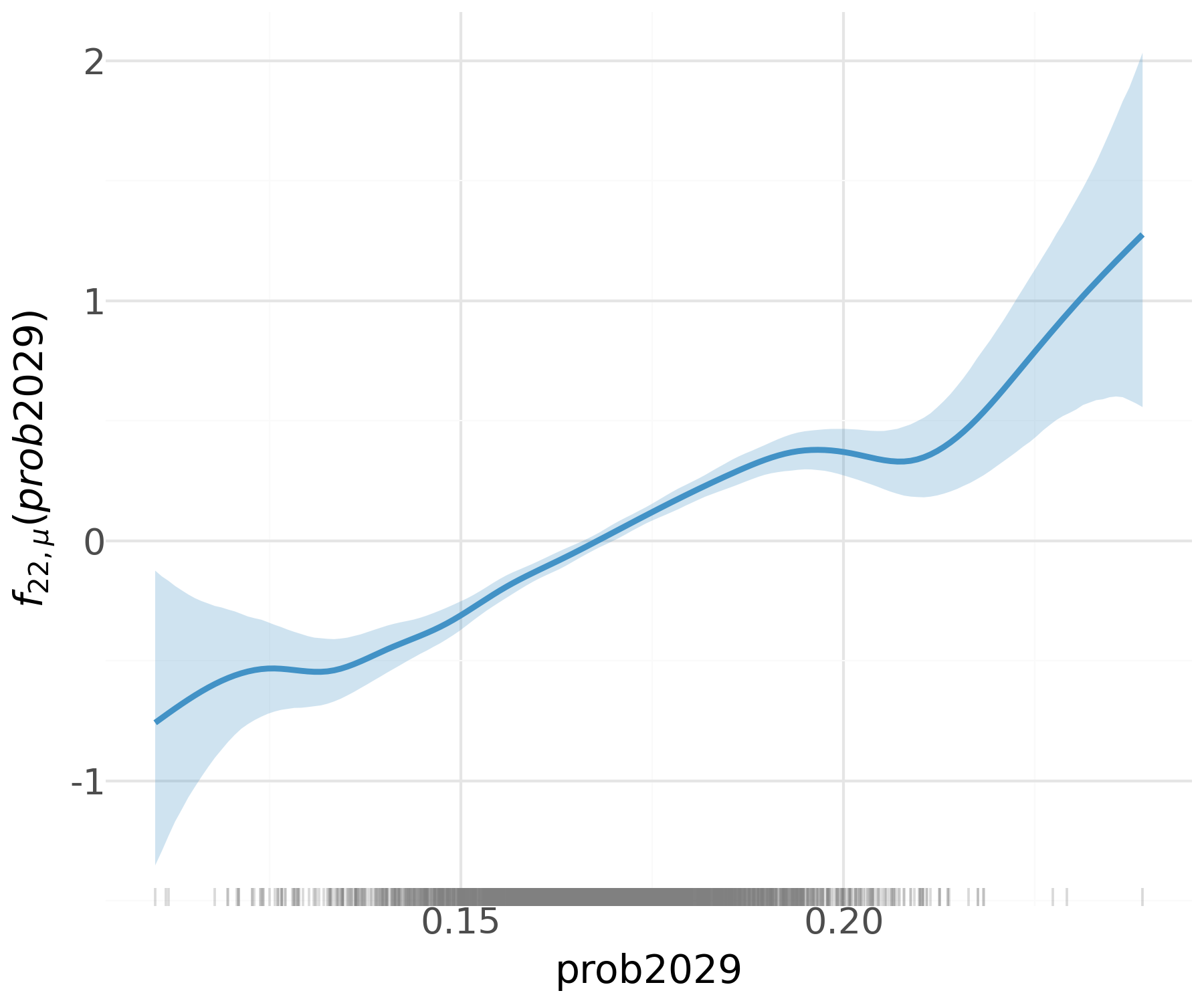}
    \end{minipage}
    \begin{minipage}{0.32\textwidth}
        \centering
        \includegraphics[width=\linewidth]{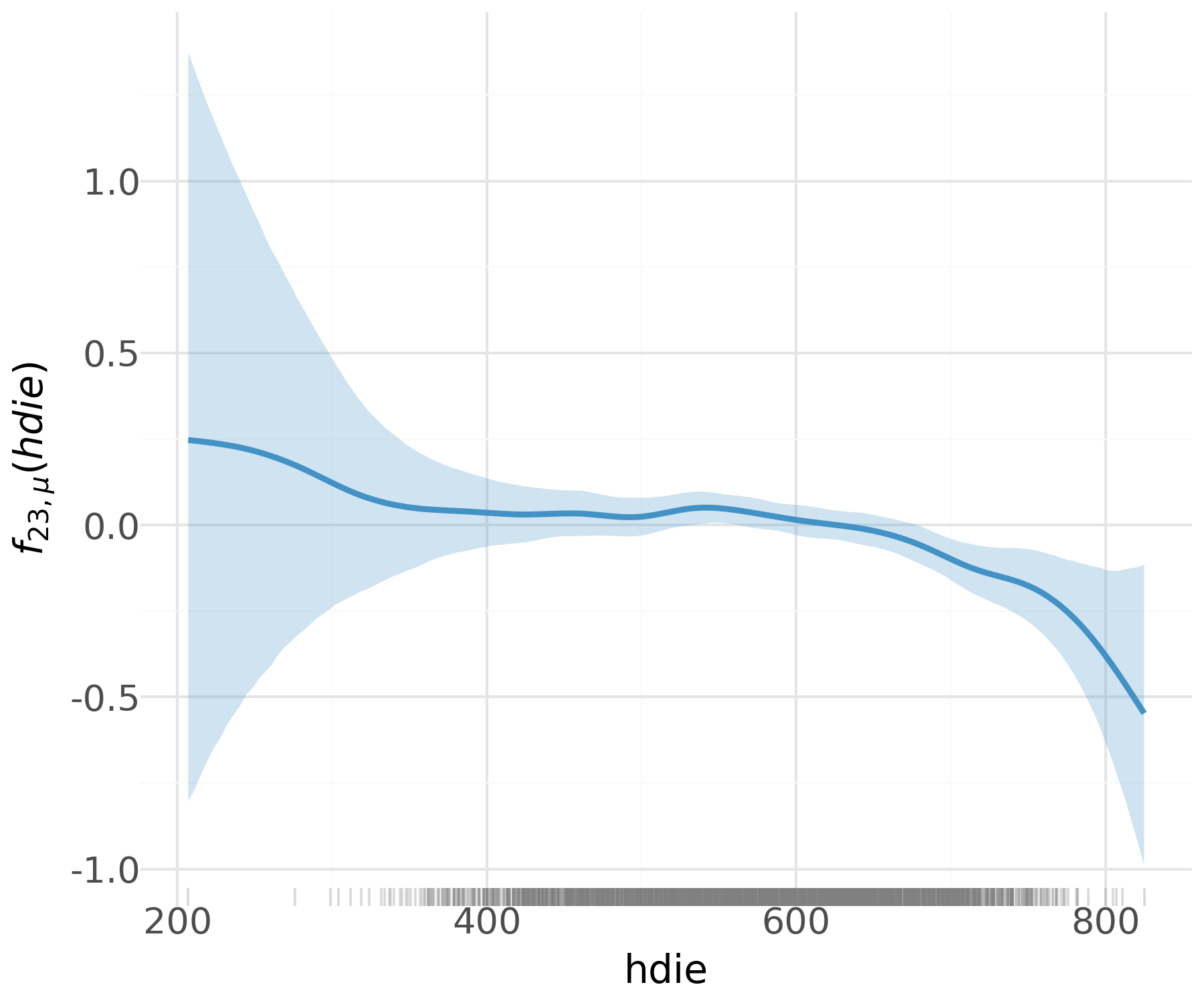}
    \end{minipage}
    \begin{minipage}{0.32\textwidth}
        \centering
        \includegraphics[width=\linewidth]{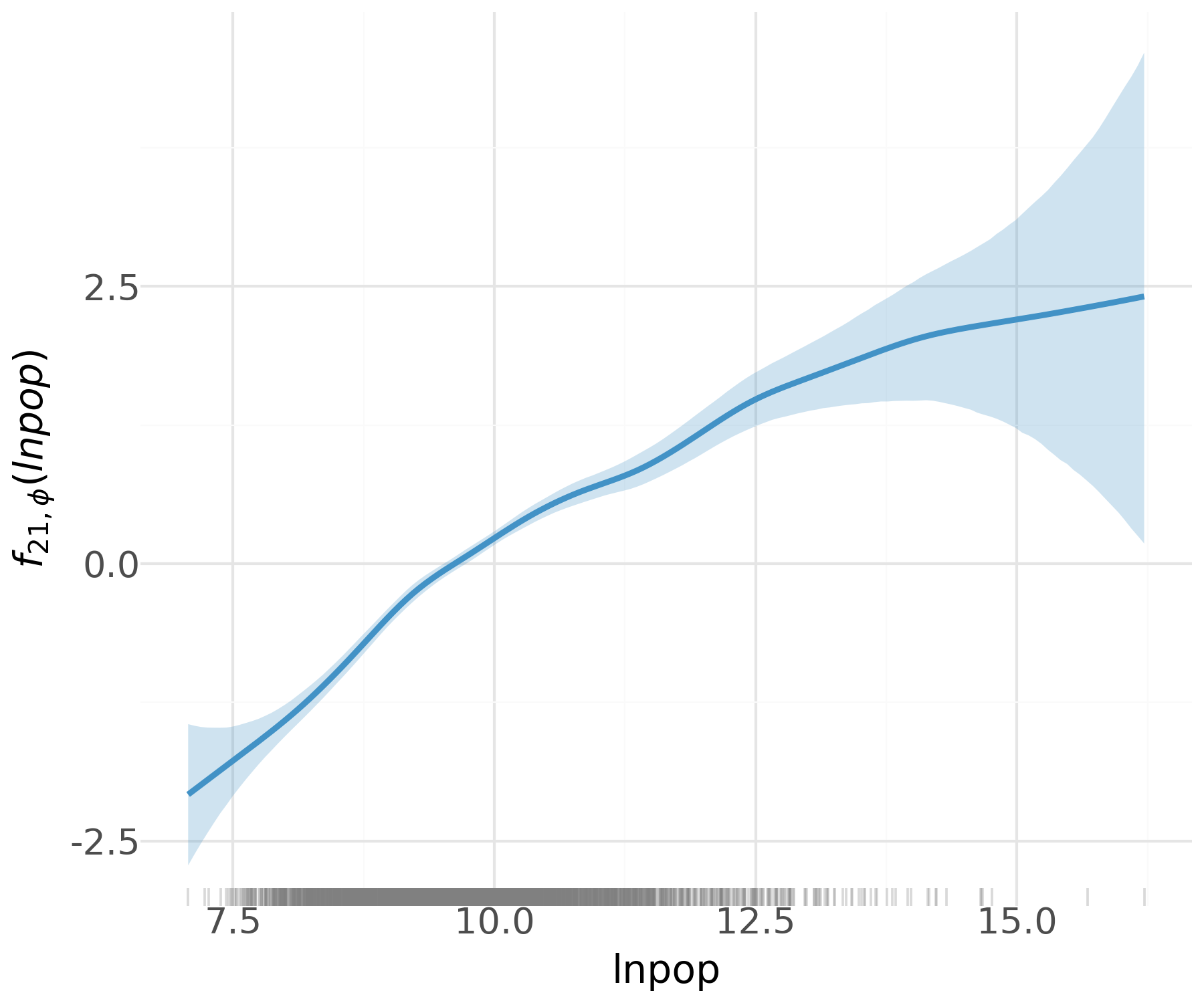}
    \end{minipage}
    \begin{minipage}{0.32\textwidth}
        \centering
        \includegraphics[width=\linewidth]{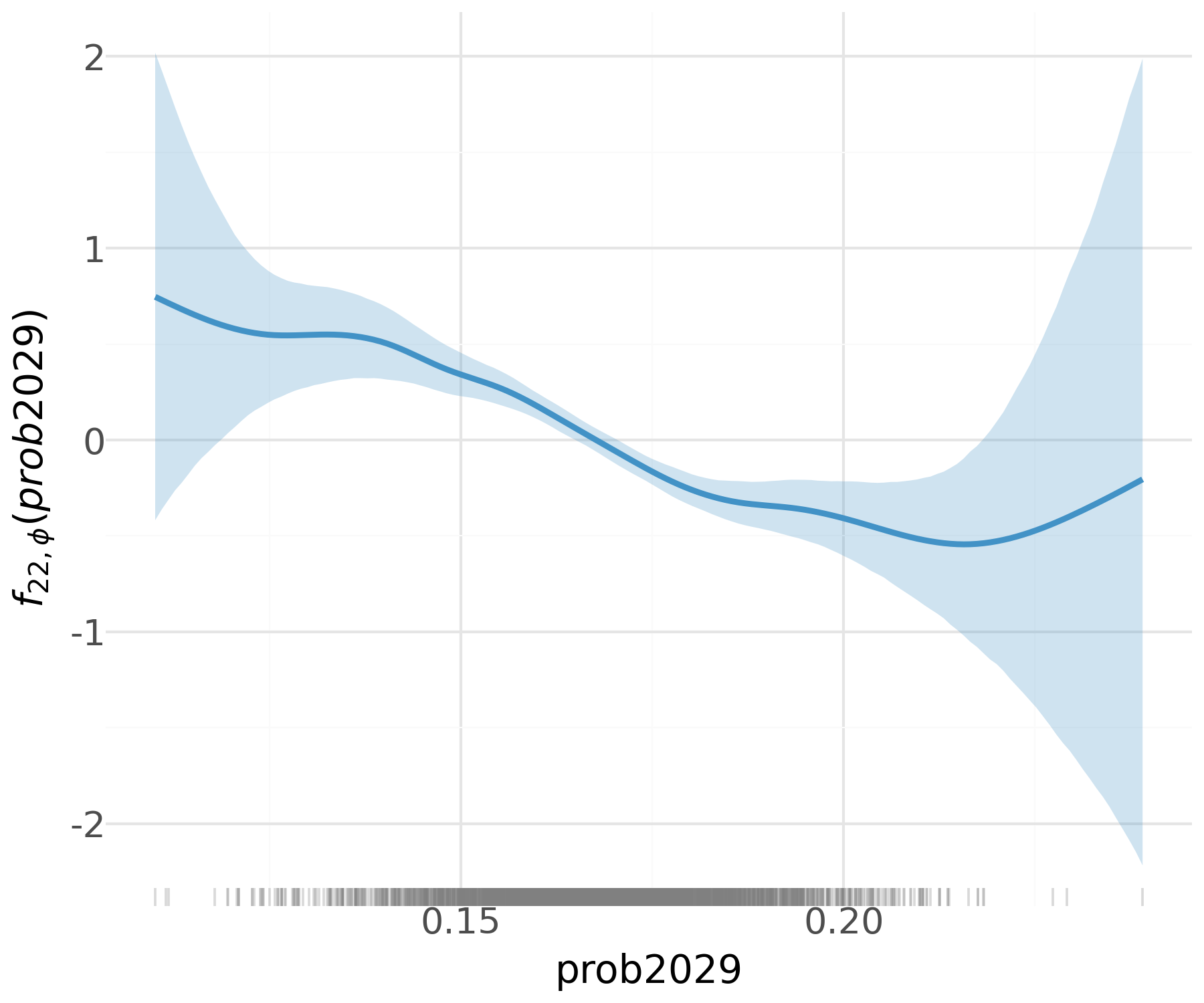}
    \end{minipage}
    \begin{minipage}{0.32\textwidth}
        \centering
        \includegraphics[width=\linewidth]{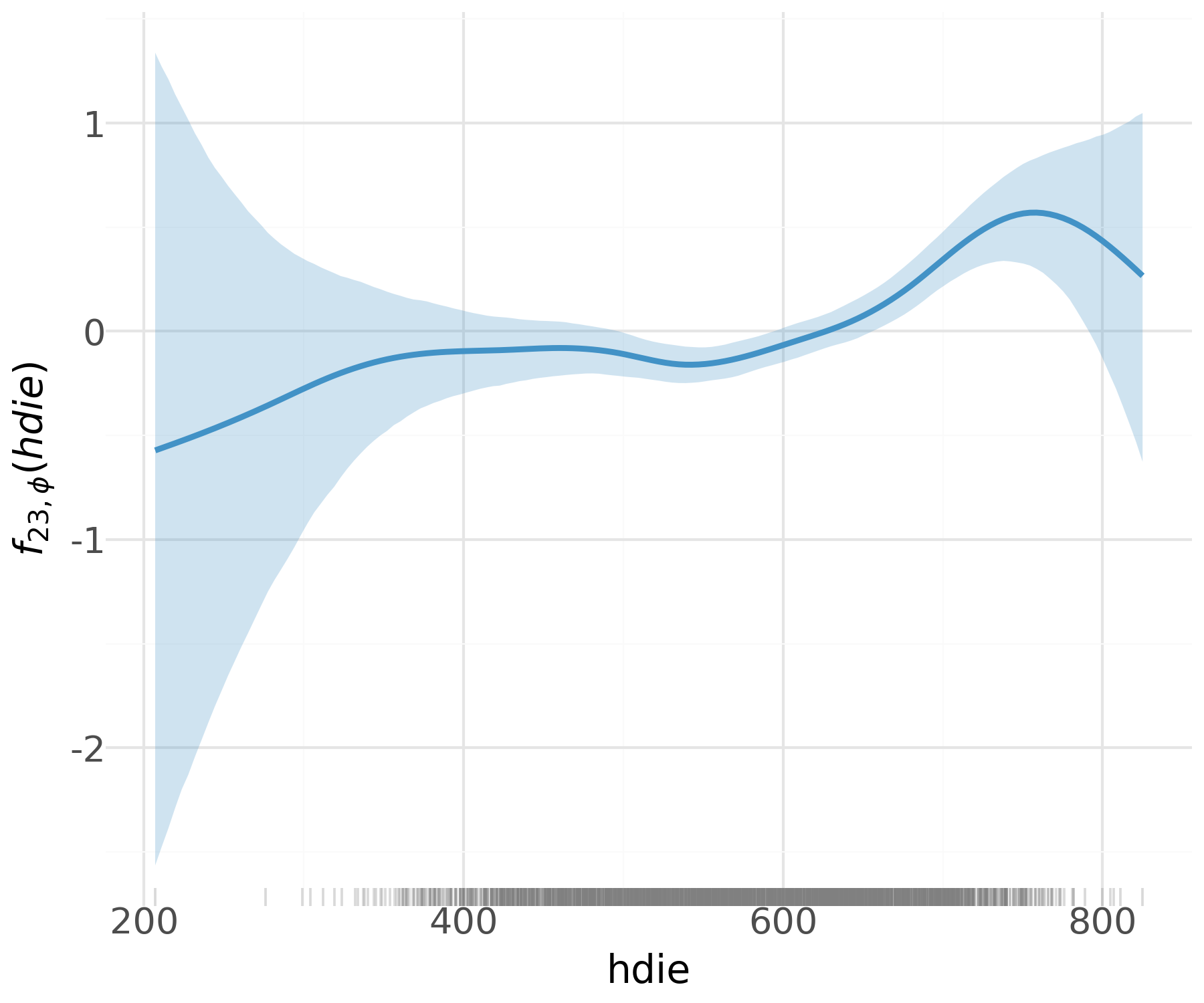}
    \end{minipage}
    \begin{minipage}{0.6\textwidth}
        \centering
        \includegraphics[width=\linewidth]{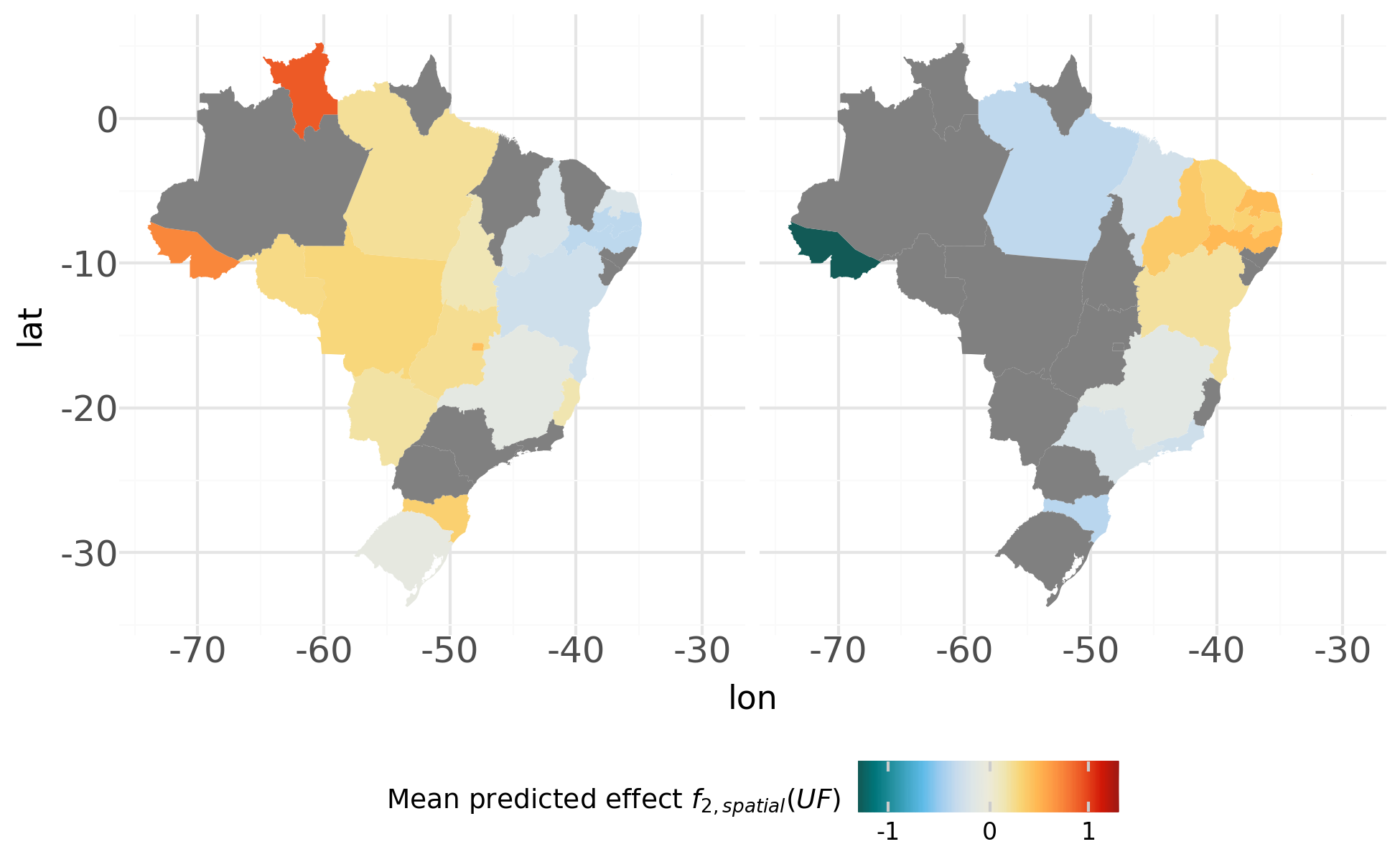}
    \end{minipage}
    \caption{Plots of the estimated effects of the continuous part of the zero-inflated beta regression model. The top row shows the estimated nonlinear effects of the conditional mean submodel, the middle row those of the conditional precision submodel, and the bottom row shows the estimated spatial effects. States whose 90\% credible intervals for the spatial effects include zero (nonsignificant) are shown in gray.}
    \label{fig:effects-cont-beta}
\end{figure}

Since the discrete components are identical in both models, we only present the results for the continuous component. Figure \ref{fig:effects-cont-beta} presents the estimated nonlinear and spatial effects for both the conditional mean and precision submodels. The estimated effects for the conditional mean in the zero-inflated beta regression closely resemble those obtained from the zero-inflated structured additive quantile regression at the median quantile. For the conditional precision parameter, the estimated effect of \texttt{lnpop} is approximately linear and positive, indicating that larger municipalities tend to have higher precision. The effect of \texttt{prop2029} is slightly decreasing, suggesting a small reduction in precision as the share of young residents increases. For \texttt{hdie}, the effect is close to zero for moderate values, but displays a convex pattern at higher values, suggesting a modest increase in the conditional precision in municipalities with high educational development. An analysis of the spatial effects in the conditional precision submodel reveals that, in general, negative effects occur in parts of the Southern, Southeastern, and Northern regions, suggesting lower precision and greater variability in those areas. Positive effects are concentrated in much of the Northeast, indicating higher precision and lower variability. Notably, Acre exhibits a strong negative effect, highlighting that municipalities in this state experience more variable traffic mortality patterns.

\begin{figure}[!ht]
    \centering
    \begin{minipage}{0.32\textwidth}
        \centering
        \includegraphics[width=\linewidth]{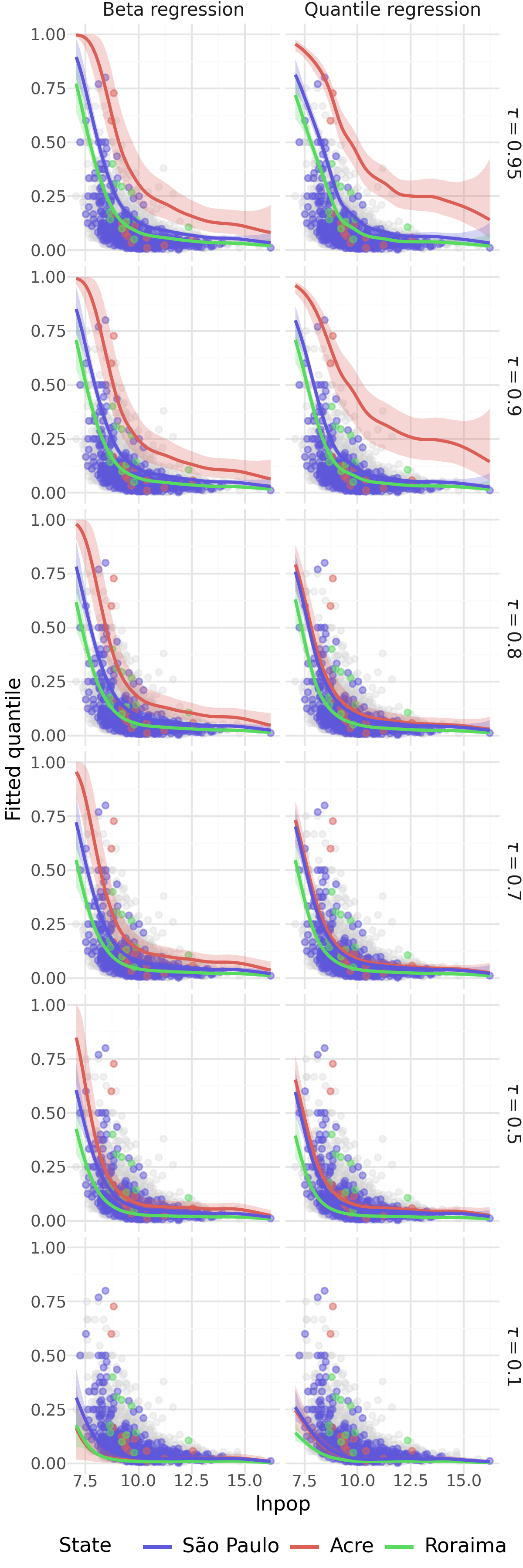}
    \end{minipage}
    \begin{minipage}{0.32\textwidth}
        \centering
        \includegraphics[width=\linewidth]{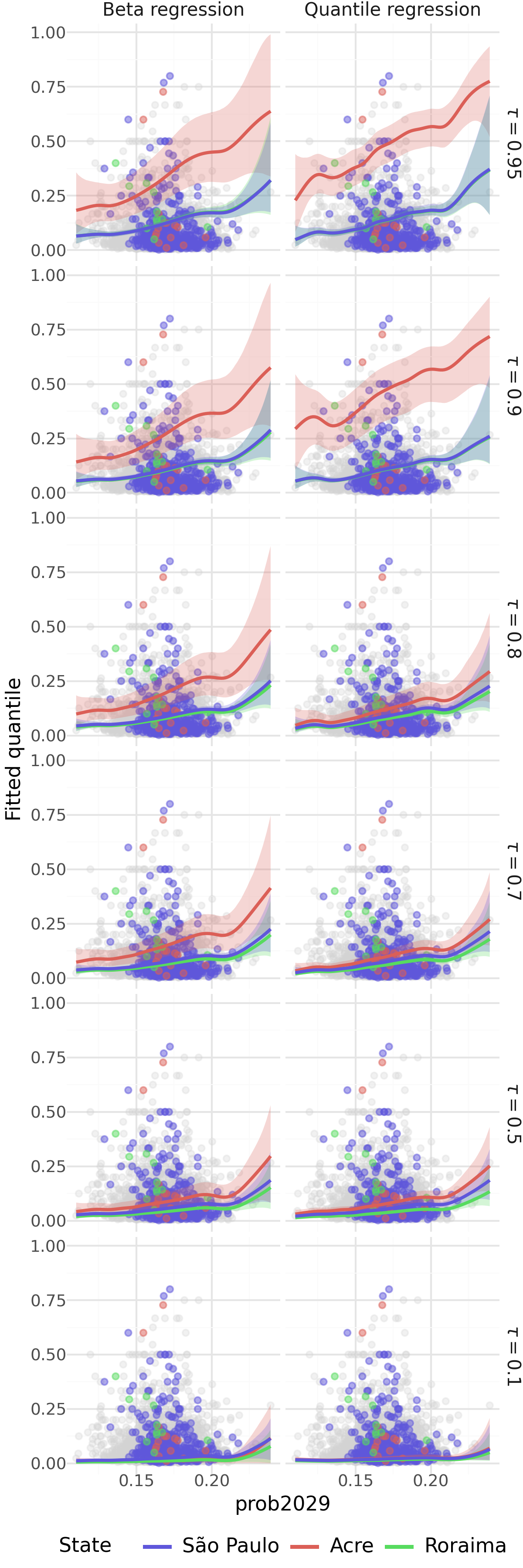}
    \end{minipage}
    \begin{minipage}{0.32\textwidth}
        \centering
        \includegraphics[width=\linewidth]{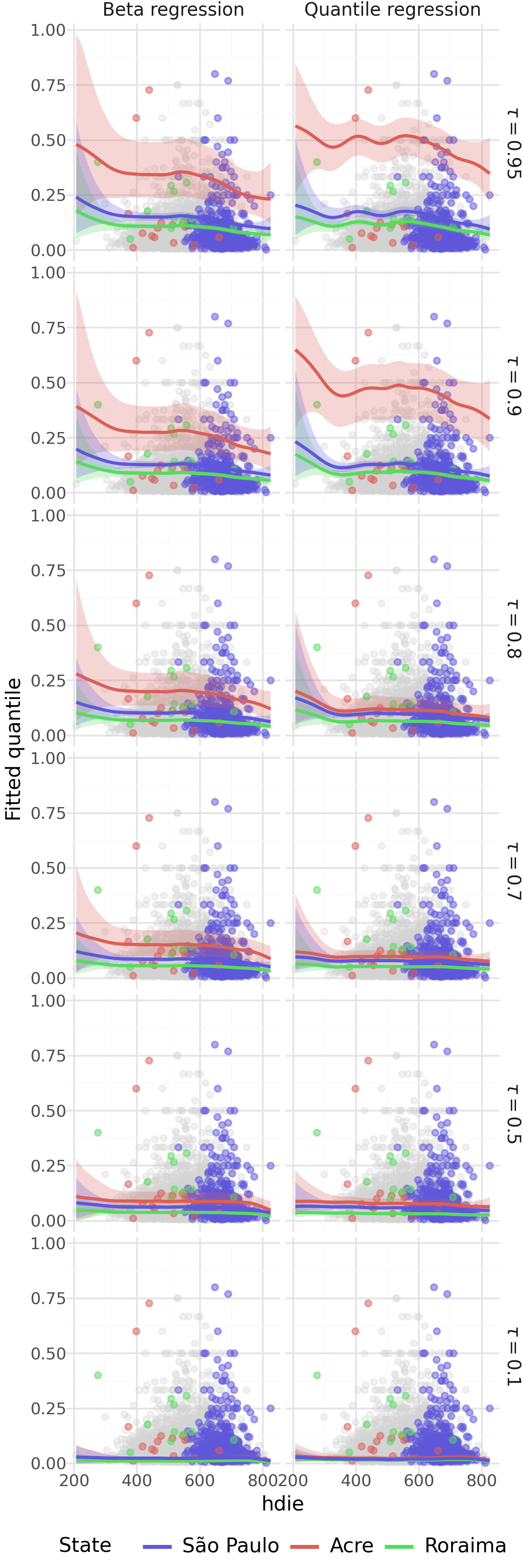}
    \end{minipage}
    \caption{Fitted quantiles of the continuous part based on the zero-inflated structured additive quantile regression and zero-inflated beta regression models. The solid lines represent the posterior mean of the predicted $\tau$th quantile for the states of Acre, Roraima and S\~ao Paulo. All remaining covariates were held fixed at their state-specific sample means. Shaded ribbons represent the $90\%$ credible intervals.}
    \label{fig:quantiles-cont}
\end{figure}

Finally, Figure \ref{fig:quantiles-cont} shows the fitted conditional quantiles for the $10\%$, $50\%$, $70\%$, $80\%$, $90\%$, and $95\%$ quantiles based on the zero-inflated structured additive quantile regression and zero-inflated beta regression models. Across all quantiles and for both models, the fitted quantiles decrease with \texttt{lnpop}, increase with \texttt{prop2029}, and exhibit a slightly decreasing relationship with \texttt{hdie}. The lower quantiles ($\tau=0.1$) are generally very close to zero and are nearly indistinguishable between the two modeling approaches. This pattern is observed in all states. For S\~ao Paulo and Roraima, the fitted quantiles are very similar across the zero-inflated structured additive quantile and the zero-inflated beta regression models, with little visible difference between the two models' estimates at all quantile levels. In contrast, for Acre, the fitted quantiles differ more markedly across models. In particular, for intermediate quantiles ($\tau=0.5$, $0.7$, and $0.8$), the zero-inflated beta regression tends to produce systematically higher fitted quantiles than the quantile regression, appearing as an upward shift of the estimated curves. This discrepancy becomes even more pronounced at the upper quantiles ($\tau = 0.9$ and $0.95$), where the fitted curves from the two models differ more clearly. At these levels, the quantile regression yields estimates with more consistent behavior and comparatively more concentrated credible intervals, whereas the beta regression shows increased uncertainty in the upper tail, especially for \texttt{prop2029} and \texttt{hdie}.

In general, the zero-inflated beta regression and the zero-inflated structured additive quantile regression yield broadly consistent conclusions regarding the direction of the covariate effects on the proportion of traffic mortality. However, the quantile regression provides additional insight into how these effects vary across different parts of the conditional distributrion. In particular, the negative association between population size and the proportion of traffic mortality and the postivie association with the proportion of young residents are more pronounced at higher quantiles, highlighting stronger effects among municipalities with relatively high proportion of traffic mortality. These distributional differences cannot be directly captured by the beta regression, which focuses on the conditional mean. While the zero-inflated beta regression remains a useful tool, the zero-inflated quantile regression offers a more detailed characterization of the covariate effects accross the full conditional distribution of the proportion of traffic mortality.

\subsection{Speech intelligibility tests data}\label{sec:application2}

We analyze data presented by \textcite{Hu2015} from a speech intelligibility experiment involving seven recipients of cochlear implants, collected to evaluate differences in speech recognition performance under various sound-processing algorithms. These data were also analyzed by \textcite[Chapter 9]{stasinopoulos2024gamlss}. In the experiment, subjects listened to a prerecorded sentence presented in the presence of background noise (an interfering talker), with the signal-to-noise ratio (SNR) varied across trials, and were asked to repeat it. Data were collected in tracks of 20 sentences each, with listening conditions held constant within a track. In this study, each sentence corresponds to a single observation, and the response variable is the proportion of morphemes correctly identified, defined as $y = w/N$, where $w$ denotes the number of morphemes correctly recognized and $N$ is the total number of morphemes in the sentence. 

The dataset includes the following covariates: the sound-processing algorithm used in the cochlear implant (\texttt{algorithm}: A, B, or C), the gender of the interfering talker (\texttt{noise\_g}: male or female), and the subject identifier (\texttt{subject}, which allows for the inclusion of subject-level random effects in the analysis). Each subject completed four tracks for each experimental conditions, resulting in $80$ sentences per condition.

A notable feature of these data is the high frequency of extreme responses: many sentences are either completely recognized ($y=1$) or completely missed ($y=0$), with approximately $34\%$ of sentences at each extreme. This results in nearly $68\%$ of observations lying at the boundaries of the unit interval $[0,1]$. \textcite{Hu2015} analyzed these data using the zero- and one-inflated beta regression model, while \textcite[Chapter 9]{stasinopoulos2024gamlss} used the zero- and one-inflated simplex regression model. In this work, we adopt the zero- and one-inflated structured additive quantile regression model with the following specifications for the predictors
\begin{small}
    \begin{align*}
        \log\left(\dfrac{p_{0{ij}}}{p_{2{ij}}}\right) & = \beta_{00} + \beta_{11} \texttt{snr}_{ij} +  \beta_{02} \texttt{algorithmB}_{ij} + \beta_{03} \texttt{algorithmC}_{ij} + \beta_{04} \texttt{noise\_gM}_{ij}\\
        & \hspace{0.3cm} + \beta_{05} \texttt{snr}_{ij} \times \texttt{algorithmB}_{ij} + \beta_{06} \texttt{snr}_{ij} \times \texttt{algorithmC}_{ij} + u_{0j},\\
        \log\left(\dfrac{p_{1{ij}}}{p_{2{ij}}}\right) & = \beta_{10} + \beta_{11} \texttt{snr}_{ij} +  \beta_{12} \texttt{algorithmB}_{ij} + \beta_{13} \texttt{algorithmC}_{ij} + \beta_{14} \texttt{noise\_gM}_{ij}\\
        & \hspace{0.3cm}  + \beta_{15} \texttt{snr}_{ij} \times \texttt{algorithmB}_{ij}  + \beta_{16} \texttt{snr}_{ij} \times \texttt{algorithmC}_{ij} + u_{1j},\\
        \text{logit}\left[Q_{y_{ij}\mid y_{ij} \in (0,1)}(\tau \mid \mathbf{x}_{ij})\right] &= \beta_{20} + \beta_{21} \texttt{snr}_{ij} +  \beta_{22} \texttt{algorithmB}_{ij} + \beta_{23} \texttt{algorithmC}_{ij} + \beta_{24} \texttt{noise\_gM}_{ij}\\
        & \hspace{0.3cm}  + \beta_{25} \texttt{snr}_{ij} \times \texttt{algorithmB}_{ij}   + \beta_{26} \texttt{snr}_{ij} \times \texttt{algorithmC}_{ij} + u_{2j},
    \end{align*}
\end{small}
where $p_{0{ij}} = \mathbb{P}(y_{ij}=0)$ and $p_{1{ij}} = \mathbb{P}(y_{ij}=1)$ are the probability of observing zero and one, respectively, $p_{2{ij}} = 1- p_{0{ij}} - p_{1{ij}}$, $Q_{y_{ij}\mid y_{ij} \in (0,1)}(\tau \mid \mathbf{x}_{ij})$ is the $\tau$th quantile of the conditional distribution of $y_{ij}$ given that $y_{ij} \in (0,1)$, for the $i$th  sentence and $j$th subject, $\beta_{00}$, $\beta_{01}$, and $\beta_{02}$ are the intercepts, and the remaining coefficients correspond to linear effects of the covariates \texttt{snr}, \texttt{algorithmB}, \texttt{algorithmC}, \texttt{noise\_gM}, and interactions. The specification of the predictors are based on the variable selection conducted by \textcite[Chapter 9]{stasinopoulos2024gamlss}. Between-subject variability is captured through the random intercepts $u_{0j}$, $u_{1j}$, $u_{2j}$, modeled as independent Gaussian terms with variance parameters assigned weak inverse-gamma priors with concentration $1$ and scale $0.005$. We consider three conditional quantiles ($10\%$, $50\%$, $90\%$) in order to capture the central tendency and overall distribution of recognition scores. In addition, the zero- and one-inflated structure models the log-odds of completely missed ($y=0$) and fully correct ($y=1$) sentences relative to the continuous response.

Table \ref{tab:summary_tau} presents the posterior means and standard deviations of the model parameters, along with the corresponding $90\%$ credible interval. For modeling purposes, all categorical variables are coded as $-1/1$ dummy variables. The sound-processing algorithms are represented by \texttt{algorithmB} and \texttt{algorithmC}, with $1$ indicating the presence of the corresponding algorithm and $-1$ indicating the reference algorithm (A). Similarly, \texttt{noise\_gM} is coded as $1$ for a male talker and $-1$ for a female talker. Under this coding scheme, each coefficient reflects half of the total effect of changing from one category to the other. Table \ref{tab:summary_tau} also reports the posterior means and standard deviations of the subject-level random intercept variances. The results for the submodels of the discrete part are similar to those obtained in \textcite[Chapter 9]{stasinopoulos2024gamlss}, and the continuous part at the median ($\tau=0.5$) are also comparable to their conditional mean submodel results, with slight differences likely due to robustness.

\begin{table}[!ht]
\centering
\caption{Posterior summaries for the three submodels: the quantile submodel at $\tau = 0.1, 0.5,$ and $0.9$, and the discrete submodels $\log(p_{0{ij}}/p_{2{ij}})$ and $\log(p_{1{ij}}/p_{2{ij}})$. For each parameter, the posterior mean, standard deviation, and $90\%$ credible interval are reported. The last row reports the posterior means and standard deviations of the subject-level random intercept variances.}
\label{tab:summary_tau}
\footnotesize
\setlength{\tabcolsep}{3pt}
\begin{tabular}{l@{}rrrrrrrrr}
\toprule
& \multicolumn{9}{c}{$\text{logit}[Q_{y_{ij}\mid y_{ij} \in (0,1)}(\tau \mid \mathbf{x}_{ij})]$} \\
\cmidrule(rr){2-10}
& \multicolumn{3}{c}{$\tau=0.1$} & \multicolumn{3}{c}{$\tau=0.5$} & \multicolumn{3}{c}{$\tau=0.9$} \\
\cmidrule(rr){2-4} \cmidrule(lr){5-7} \cmidrule(lr){8-10}
       & Mean & SD  &       $90\%$ CI         & Mean & SD  &   $90\%$ CI   & Mean & SD  &  $90\%$ CI  \\
\midrule
\texttt{Intercept}   & $-1.646$ & $0.040$ & $(-1.713,~-1.583)$ & $-0.308$ & $0.060$ & $(-0.407,~-0.211)$ & $1.505$  & $0.045$ & $(1.428,~1.576)$ \\
\texttt{snr}            & $0.075$  & $0.010$ & $(0.059,~0.091)$   & $0.095$  & $0.011$ & $(0.076,~0.113)$  & $0.027$  & $0.008$ & $(0.014,~0.040)$ \\
\texttt{algorithmB}     &  $0.026$  & $0.035$ & $(-0.032,~0.082)$  & $0.067$  & $0.057$ & $(-0.025,~0.161)$  & $0.022$  & $0.035$ & $(-0.036,~0.080)$ \\
\texttt{algorithmC}     & $0.011$  & $0.033$ & $(-0.043,~0.064)$  & $0.017$  & $0.050$ & $(-0.064,~0.098)$  & $-0.017$ & $0.042$ & $(-0.088,~0.052)$ \\
\texttt{noise\_gM}  & $0.041$  & $0.022$ & $(0.005,~0.076)$   & $0.051$  & $0.026$ & $(0.009,~0.094)$  & $0.033$  & $0.015$ & $(0.008,~0.059)$ \\
\texttt{snr}:\texttt{algorithmB}  &$0.009$  & $0.010$ & $(-0.007,~0.026)$  & $-0.011$ & $0.013$ & $(-0.033,~0.010)$  & $0.001$  & $0.008$ & $(-0.012,~0.013)$ \\
\texttt{snr}:\texttt{algorithmC} & $-0.011$ & $0.010$ & $(-0.027,~0.005)$  & $-0.022$ & $0.012$ & $(-0.042,~-0.003)$ & $-0.001$ & $0.008$ & $(-0.014,~0.012)$ \\
\midrule
\texttt{random(subject)}      & $0.003$  & $0.003$ &  & $0.007$  & $0.007$ &  & $0.003$  & $0.002$ & \\
\midrule \addlinespace
& \multicolumn{3}{c}{$\log(p_{0{ij}}/p_{2{ij}})$} & \multicolumn{3}{c}{$\log(p_{1{ij}}/p_{2{ij}})$} &  & &  \\
\cmidrule(rr){2-4} \cmidrule(lr){5-7}
       & Mean & SD    & $90\%$ CI & Mean & SD    & $90\%$ CI &  & & \\
\cmidrule(ll){1-7}
\texttt{Intercept}      & $0.633$  & $0.089$ & $(0.488,~ 0.776)$  & $-0.909$ & $0.143$ & $(-1.140,~ -0.677)$ &  & & \\
\texttt{snr}            & $-0.221$ & $0.015$ & $(-0.245,~ -0.197)$ & $0.213 $ & $0.014$ & $(0.190,~ 0.237)$&  & & \\
\texttt{algorithmB}     & $-0.130$ & $0.058$ & $(-0.226,~ -0.034)$ &  $0.060 $ & $0.073$ & $(-0.060,~ 0.181)$&   & & \\
\texttt{algorithmC}     & $-0.174$ & $0.060$ & $(-0.273,~ -0.074)$ & $-0.131$ & $0.075$ & $(-0.254,~ -0.007)$ &  & & \\
\texttt{noise\_gM}  & $-0.161$ & $0.032$ & $(-0.214,~ -0.108)$ & $0.151 $ & $0.033$ & $(0.098,~ 0.205)$&  & & \\
\texttt{snr}:\texttt{algorithmB} & $0.050$  & $0.016$ & $(0.024,~ 0.075)$  & $-0.027$ & $0.015$ & $(-0.051,~ -0.002)$ &  & & \\
\texttt{snr}:\texttt{algorithmC} &  $0.011$  & $0.016$ & $(-0.016,~ 0.038)$  & $-0.014$ & $0.015$ & $(-0.039,~ 0.011)$ &  & & \\
\cmidrule(ll){1-7}
\texttt{random(subject)}        & $0.036$  & $0.030$ &  & $0.118 $ & $0.088$ & &  & & \\
\bottomrule
\end{tabular}
\end{table}

The conditional quantile results show how experimental conditions are associated with different levels of speech recognition performance in cochlear implant users, focusing on sentences that are partially understood, i.e., excluding fully correct and fully missed responses. From Table \ref{tab:summary_tau}, across all quantiles, higher signal-to-noise ratio is associated with higher values of the conditional quantiles of partially recognized morphemes, with stronger effects at lower to median quantiles ($\tau=0.1$ and $0.5$) and weaker effects at the upper quantile ($\tau=0.9$). The gender of the interfering talker also presents a small but positive effect across all quantiles. In addition, a negative interaction between SNR and Algorithm C is observed for $\tau=0.5$. For the discrete part, higher SNR is associated with lower odds of completely missed sentences and higher odds of fully correct sentences, suggesting that better listening conditions are related to a greater tendency toward partial or full recognition. Finally, male interfering talkers are associated with higher odds of partial or full recognition, and an interaction between SNR and Algorithm C is observed. 

\begin{figure}[!ht]
    \centering
    \begin{minipage}{0.4\textwidth}
        \centering
        \includegraphics[width=\linewidth]{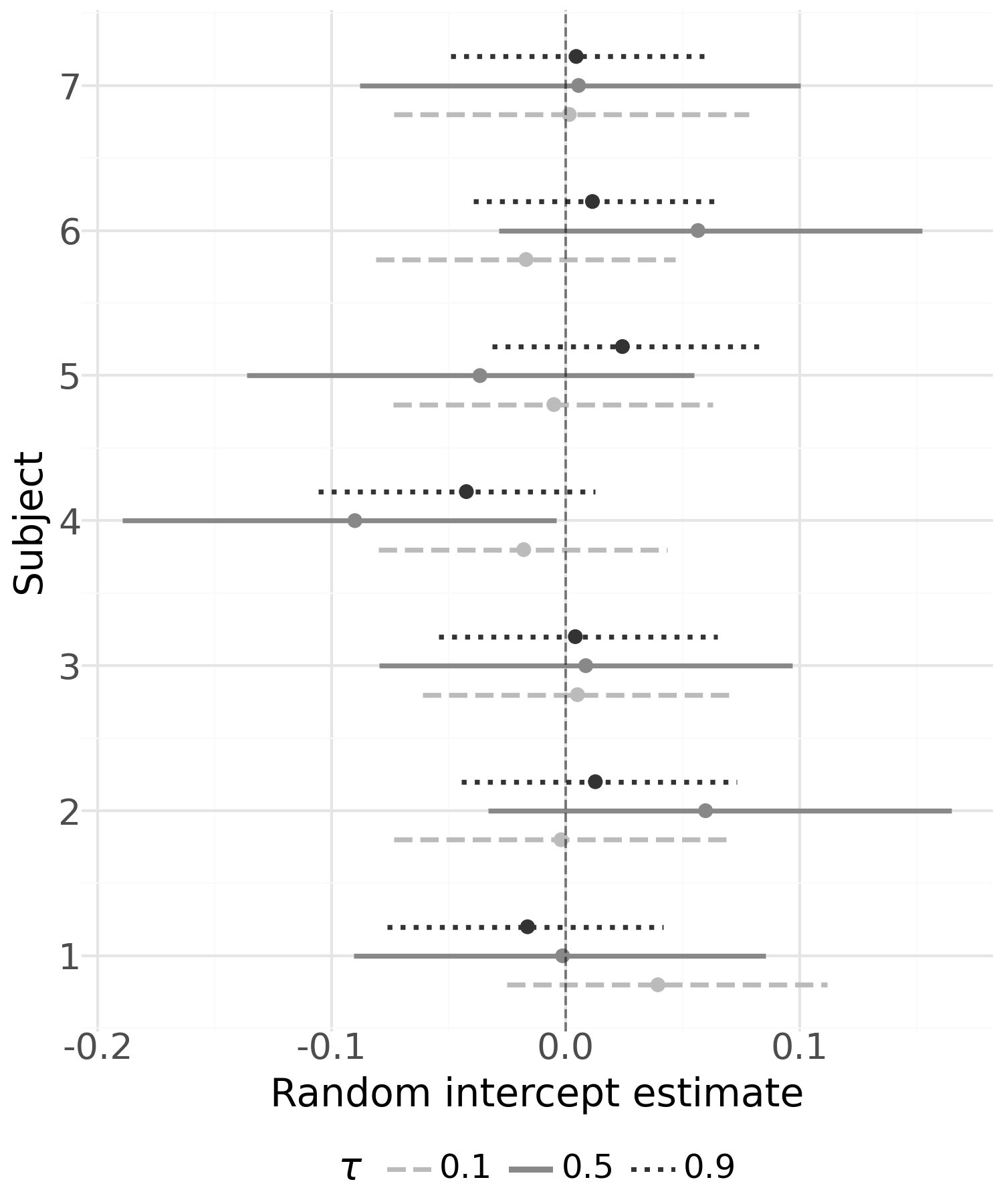}
    \end{minipage}
    \begin{minipage}{0.4\textwidth}
        \centering
        \includegraphics[width=\linewidth]{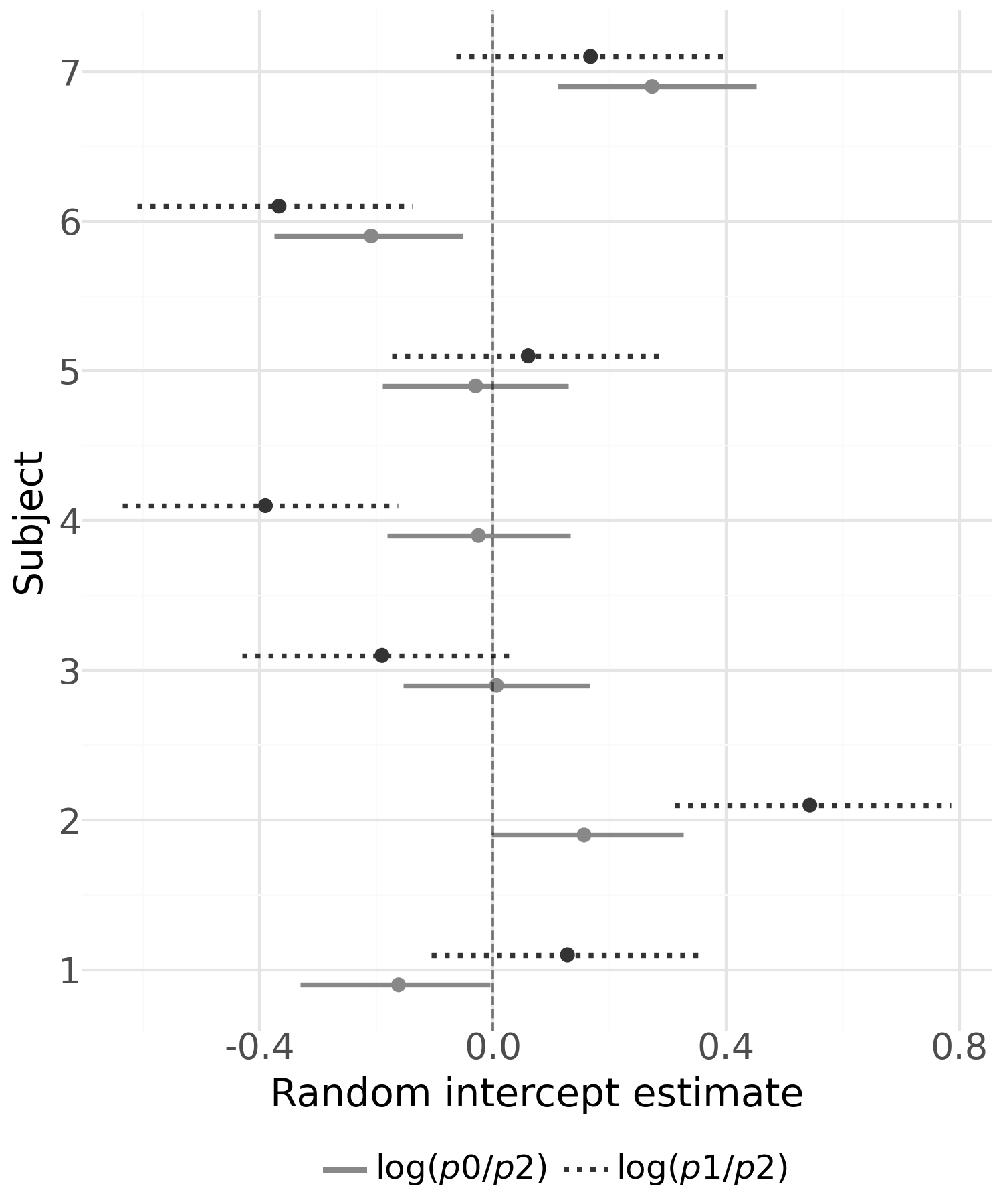}
    \end{minipage}\hfill
    \caption{Forest plot summarizing the random intercepts for each subject. The left panel shows the estimates for the continuous component across the $0.1$, $0.5$, and $0.9$ quantiles, while the right panel shows the estimates for the discrete submodels. Each horizontal line represents the posterior mean of the random intercept for a subject, with the line length indicating the $90\%$ credible interval.}
    \label{fig:randomintercept}
\end{figure}

Figure \ref{fig:randomintercept} presents the forest plots summarizing the subject-level random intercepts for both the continuous and discrete part. The plots reveal distinct patterns of between-subject variability. In the continuous component, estimated random effects are generally small, with posterior means close to zero and $90\%$ credible intervals mostly including zero, indicating limited between-subject variability. Nevertheless, they remain important to account for repeated measurements per subject. In contrast, there is greater between-subject variability in the discrete component, as reflected by the posterior variances of the subject-level random intercepts; see also Table \ref{tab:summary_tau}. Figure \ref{fig:randomintercept} shows that, for the submodels of the discrete part, several subjects show credible intervals that do not include zero, indicating differences in their propensity to produce extreme outcomes. Overall, this highlights that individual differences are most pronounced in the probability of complete failures or perfect recognitions, while intermediate performance is relatively homogeneous across subjects.

\begin{figure}[!ht]
    \centering
    \begin{minipage}{0.5\textwidth}
        \centering
        \includegraphics[width=\linewidth]{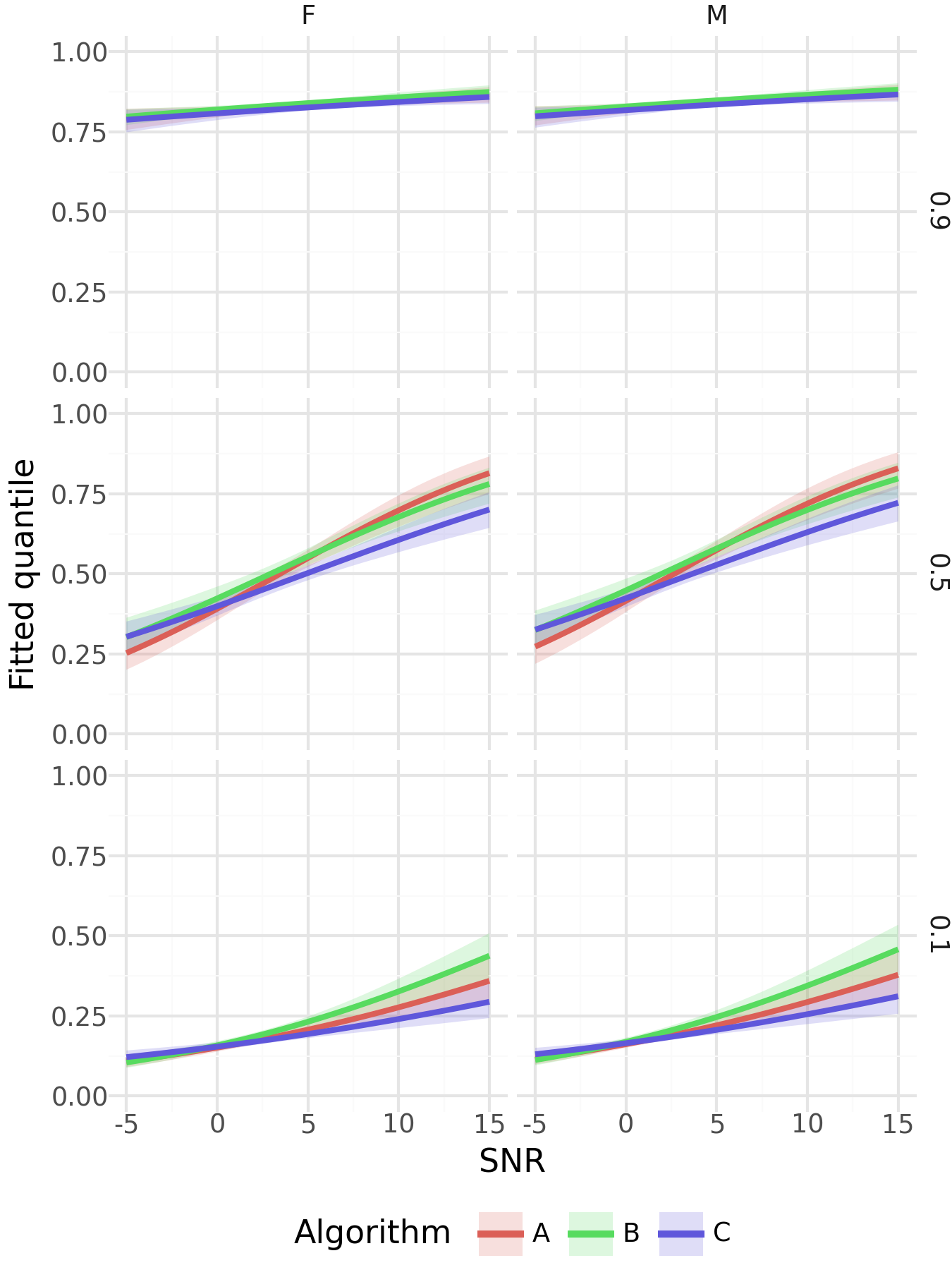}
    \end{minipage}\hfill
    \begin{minipage}{0.5\textwidth}
        \centering
        \includegraphics[width=\linewidth]{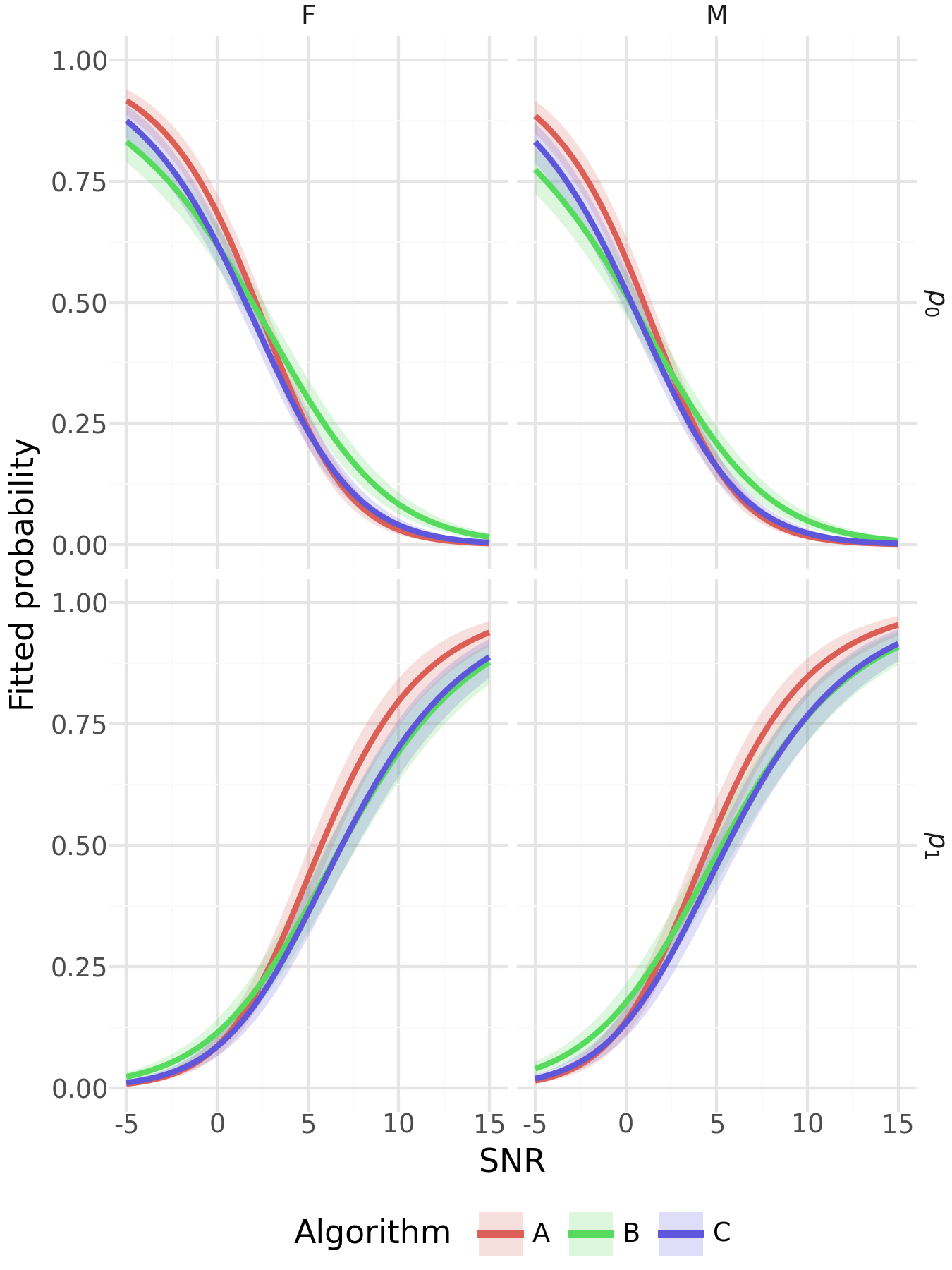}
    \end{minipage}\hfill
    \caption{Fitted quantiles (left panel) and probabilities (right panel), with 90\% credible interval, plotted against SNR for noise gender female (left-hand column) and male (right-hand column) across Algorithms A, B, and C. The left panel displays the fitted conditional quantiles of the continuous component at $\tau = 0.1$ (bottom), $0.5$ (middle), and $0.9$ (top), while the right panel shows the fitted probabilities of zero ($p_0$, top) and one ($p_1$, bottom).}
    \label{fig:fittedcurves}
\end{figure}

To further investigate differences in performance among the three sound-processing algorithms, Figure \ref{fig:fittedcurves} presents the fitted conditional quantiles of the continuous component and the fitted probabilities of complete failures ($y=0$) and perfect recognition ($y=1$) as function of SNR, for both female and male interfering talkers. For visualization purposes, the subject-level random intercepts were set to zero. Overall, we observe that as SNR increases, the proportion of correctly recognized morphemes also tends to increase. In addition, the fitted quantile and probabilities curves show little difference between male and female interfering talkers.

For the upper quantile ($\tau=0.9$), the three algorithms exhibit very similar performance across the entire SNR range, suggesting that, for sentences with relatively higher proportions of recognition among sentences that were neither completely missed nor perfectly recognized, the choice of algorithm has little influence. For $\tau=0.5$, corresponding to the conditional median within the continuous component, Algorithms B and C yield higher fitted values than Algorithm A at low SNR, suggesting that they may facilitate recognition under more challenging listening conditions. Across the mid-range of SNR, all algorithms seem to perform similarly, suggesting that the effect of the algorithm is relatively small under moderate listening conditions. At high SNR, Algorithms A and B tend to outperform Algorithm C, suggesting that Algorithm C may be slightly less effective under more favorable listening conditions. These results for the conditional median quantile agree with those reported by \textcite[Chapter 9]{stasinopoulos2024gamlss} for the conditional mean of the zero- and one-inflated simplex regression model. For the lower quantile ($\tau=0.1$), representing sentences with the lowest proportions of recognition among sentences that were neither completely missed nor perfectly recognized, all three algorithms perform similarly at low to mid SNR. From mid to high SNR, algorithm B results in the highest fitted values, A is intermediate, and C is the lowest, suggesting that Algorithm B may provide the greatest benefit for more difficult sentences, while Algorithm C seems to remain less effective even under relatively favorable listening conditions.

Finally, from Figure \ref{fig:fittedcurves}, the fitted probabilities of completely missed sentences ($y=0$) and fully correct sentences ($y=1$) suggest differences in algorithm performance depending on SNR. For the fitted probability of zero ($p_0$), at low SNR, Algorithm A yields the highest fitted probability of a completely missed sentence, C is intermediate, and B the lowest, suggesting that B may help reduce complete failures under challenging listening conditions. In the mid-range of SNR, all algorithms tend to have similar probabilities of zeros, indicating minimal differences between them. At high SNR, Algorithms A and C have similar fitted probabilities of zeros, while B appears slightly higher, suggesting a modestly higher number of failures for B under favorable conditions. For the fitted probability of one ($p_1$), at low SNR, all algorithms have similar fitted probabilities, with B slightly higher, suggesting only minor differences when listening conditions are difficult. In the mid-range of SNR, all algorithms appear similar. From mid to high SNR, Algorithm A results in the highest fitted probability of fully correct responses, with B and C very similar, suggesting that A may be more effective at producing perfect recognition under good listening conditions. Overall, considering both the probabilities of zeros and ones, Algorithm B appears to perform better under low SNR, by reducing complete failures, whereas Algorithm A seems more favorable under high SNR, achieving slightly higher chances of fully correct sentences.

\section{Concluding remarks}\label{sec:conclusions}

In this paper, we propose a flexible quantile regression model for bounded continuous data with observations at the boundaries. The zero- or one-inflated structured additive quantile regression models extend the quantile regression model for zero- or one-inflated bounded data proposed by \textcite{Santos2015} by incorporating structured additive regression, allowing for the inclusion of spatial effects, random effects, varying-coefficient terms, and others. We also propose the zero- and one-inflated structured additive quantile regression models for analyzing bounded continuous data with a non-negligible amount of observations at both boundaries of the unit interval. We consider the two-part modeling framework, enabling separate modeling of the quantiles of the conditional continuous response and the probabilities of observing zeros and/or ones. In both models, both the conditional quantile of the response variable and the probabilities of zero and/or one are related to the predictors through link functions. Inference is performed in a Bayesian framework using the asymmetric Laplace distribution and efficient Markov chain Monte Carlo algorithms, implemented in the Liesel probabilistic programming framework. Simulation results demonstrate that the proposed models provide accurate estimation of parameters for both the discrete and continuous components, with good predictive performance and credible interval coverage.         

The application on traffic-related mortality data across Brazilian municipal districts, presented in Section~\ref{sec:application1}, highlights the need for structured additive predictors to adequately capture complex covariate effects. In particular, the results reveal pronounced nonlinear and spatial patterns, which cannot be accommodated by the zero-inflated quantile regression model introduced by \textcite{Santos2015}, as it is restricted to simple linear predictors. The use of the zero- and one-inflated structured additive quantile regression model is illustrated through the analysis of speech intelligibility data in Section~\ref{sec:application2}. In this application, the proposed model allows the inclusion of random effects to account for subject-specific variability, properly capturing the hierarchical structure of the data.

Finally, the methodology proposed here can be naturally extended to model positive continuous data with zero-inflation, i.e., defined on $[0, \infty)$. In our approach, the link function relating the conditional quantiles to the additive predictors is currently defined on $(0,1)$, since the continuous response is restricted to the unit interval; see Equation~\eqref{eq:continuous}. For zero-inflated positive continuous data, the same framework can be applied by defining the link function on $(0, \infty)$, allowing the model to accommodate a broader class of bounded or unbounded continuous responses.

\section*{Acknowledgments}
This study was financed, in part, by the S\~ao Paulo Research Foundation (FAPESP), Brazil, Process Number 2024/08343-9.

\printbibliography

\end{document}